\documentclass[fleqn,usenatbib]{mnras}

\usepackage[T1]{fontenc}

\DeclareRobustCommand{\VAN}[3]{#2}
\let\VANthebibliography\thebibliography
\def\thebibliography{\DeclareRobustCommand{\VAN}[3]{##3}\VANthebibliography}

\usepackage{graphicx}	
\usepackage{amsmath}	
\usepackage{amssymb}	

\usepackage{colortbl}
\usepackage{color}
\usepackage{xcolor}

\usepackage{newtxtext,newtxmath}
\usepackage{multicol}
\usepackage{multirow}

\title[QSOs Environment]{Nuclear activity in $z<0.3$ QSO 2's mainly triggered by galaxy mergers}

\author[B. L. C. Araujo et al.]{Bruna L. C. Araujo$^{1}$\thanks{E-mail: araujo.brunalc@gmail.com},
Thaisa Storchi-Bergmann$^{2}$,
Sandro B. Rembold$^{2,3}$,
André L. P. Kaipper$^{3}$,
\newauthor Bruno Dall'Agnol de Oliveira$^{2}$
\\
$^{1}$Departamento de Física, CCN, Universidade Federal do Piauí, 64049-550, Teresina, PI, Brazil\\
$^{2}$Departamento de Física, IF, Universidade Federal do Rio Grande do Sul, CP 15051, 91501-970, Porto Alegre, RS, Brazil\\
$^{3}$Departamento de Física, CCNE, Universidade Federal de Santa Maria, 97105-900, Santa Maria, RS, Brazil
}

\date{Accepted XXX. Received YYY; in original form ZZZ}

\pubyear{2022}

\begin{document}
\label{firstpage}
\pagerange{\pageref{firstpage}--\pageref{lastpage}}
\maketitle

\begin{abstract}
We investigate the role of the close environment on the nuclear activity of a sample of 436 nearby ($z<0.3$) QSO\,2's -- selected from SDSS-III spectra, via comparison of their environment and interaction parameters with those of a control sample of 1308 galaxies. We have used the corresponding SDSS images to obtain the number of neighbour galaxies $N$, tidal strength parameter $Q$ and asymmetry parameters. We find a small excess of $N$ in the QSOs compared to its three controls, and no difference in $Q$. The main difference is an excess of asymmetry in the QSOs hosts, which is almost twice that of the control galaxies.
This difference is not due to the hosts' morphology, since there is no difference in their Galaxy Zoo classifications. 
HST images of two highly asymmetric QSO\,2 hosts of our sample show that both sources have a close companion (at projected separations $\sim$\,5 kpc), which we thus conclude is the cause of the observed asymmetry in the lower resolution SDSS images.
The mean projected radius of the controls is $ \langle r \rangle = 8.53\pm$0.06\,kpc, while that of the QSO hosts is $ \langle r \rangle  = 9.39\pm$0.12\,kpc, supporting the presence of interaction signatures in the outer regions of the QSO hosts. Our results favour
a scenario in which nuclear activity in QSO\,2's is triggered by close galaxy interactions -- when the distance between the host and companion is of the order of the galaxy radius, implying that they are already in the process of merger.
\end{abstract}




\section{Introduction}\label{sec: intro}

Nuclear activity in galaxies -- which occurs when a central supermassive black hole (SMBH) is powered by the capture of matter -- is a fundamental process in the evolution of galaxies, as it triggers a series of feedback processes. These processes are observed in Active Galactic Nuclei (AGN) where they seem to impact the host galaxy and its evolution \citep{2012ARA&A..50..455F, 2014ARA&A..52..589H, 2018NatAs...2..198H}. 
But how is the nuclear activity triggered? This triggering can occur via several processes \citep{2019NatAs...3...48S}, and, at the largest scales ($>10\,$kpc), galaxy interactions can  drive a flux of matter to the central region of a galaxy that will ultimately feed the SMBH \citep{2006MNRAS.369.1808C, 2008ApJS..175..356H, 2015MNRAS.447.2123C} and trigger an AGN. Looking at the broader environment, galactic clusters could supply feeding material for AGNs by the intake of cold gas streams (cooling flows) from the hot intergalactic medium \citep{2009ApJ...702...63W, 2014ApJ...789..153L, 2016Natur.534..218T}. In addition, dense environments in galaxy clusters can exert enough pressure on the gas within galaxies to produce what are called jellyfish galaxies,
where ram pressure stripping can provide a source of AGN feeding
\citep{2017Natur.548..304P}.

Studies on the relation between the galaxy environment and nuclear activity have, however, been controversial. \citet{2003ApJ...597..142M} have found no variations in the fraction of galaxies hosting an optically-detected AGN in the SDSS survey as a function of the local environmental density. \citet{2009ApJ...695..171S} have found a similar result for a sample of X-ray selected AGN, with no difference between AGN fractions in the field and in clusters. At very small scales, \citet{2021ApJ...923....6J} have found that, both in pairs and ongoing mergers, the fraction of galaxies hosting an optical AGN are indistinguishable from that in isolated galaxies. On the other hand, \citet{2006ApJ...643...68S} have found an excess of close neighbours around quasars at scales smaller than 0.5 Mpc.
\citet{2006MNRAS.373..457L} and \citet{2019MNRAS.487.2491E} have also shown that there is an excess of incidence of nuclear activity in galaxy pairs when compared to galaxies that are not in pairs. Other studies have shown a clear relation between mergers and the triggering of AGNs \citep{2011MNRAS.418.2043E, 2011MNRAS.410.1550R, 2012MNRAS.426..276B, 2015ApJ...806..218G, 2016ApJ...822L..32F, 2020MNRAS.499.2380S, 2023ApJ...942..107S}.

In a detailed study of nine type 2 QSO host galaxies (hereafter QSO 2's or just QSOs) using HST broad and narrow-band images, selected for the study due to their hight luminosity \cite{2018ApJ...868...14S} have shown that in at least six of them there are clear signatures of interaction with a neighbour galaxy, which indicates a possible correlation between the close environment and the nuclear activity in this sample. 

In order to investigate this issue further, we have selected a sample of 436 QSO\,2's at z $\le$ 0.3 (drawn from the QSO's sample of \citet{2008AJ....136.2373R}) and a control sample of 1308 galaxies to quantify the effect of the environment on the triggering of luminous AGN. We thus present a statistical study of this sample, obtaining the number of neighbours, the tidal strength parameter $Q$, the galaxy asymmetries and morphological classification, and comparing the results for the QSOs with those of a control sample. 

This paper is structured as follows. In section \ref{sec: data}, we describe the selection of the galaxies in the QSO and control samples. In section \ref{sec: method}, we present the methodology for calculating the parameters. In section \ref{sec:result} we show our results. In section \ref{sec: discuss} we discuss the results, and in section \ref{sec: concl} we present our conclusions.
Throughout this paper we use a Hubble constant of  $H_0=70$\,km\,s$^{-1}$\,Mpc$^{-1}$ and cosmological density parameters
$\Omega_m=0.3$ and $\Omega_\Lambda=0.7$.

\section{Sample and Data}\label{sec: data}

In this section we discuss how we selected the QSO and matched control sample of galaxies for the study of their environment, which was made via the use of SDSS-III images and spectra \citep{2011AJ....142...72E} and SDSS-IV spectra \citep{2017AJ....154...28B}.

\subsection{AGN sample}

\begin{figure*}
    \centering
    \includegraphics[width=8cm]{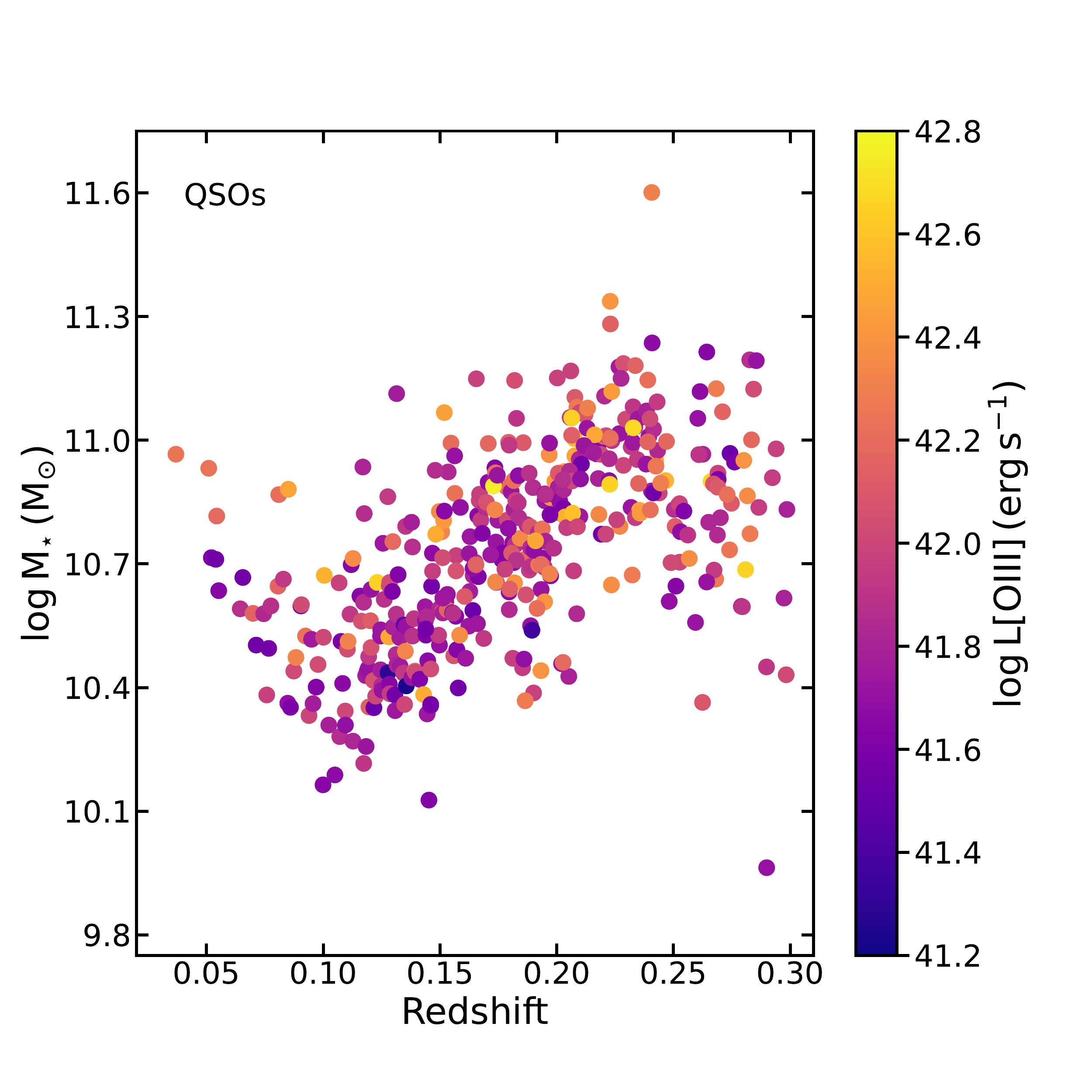}
    \includegraphics[width=8cm]{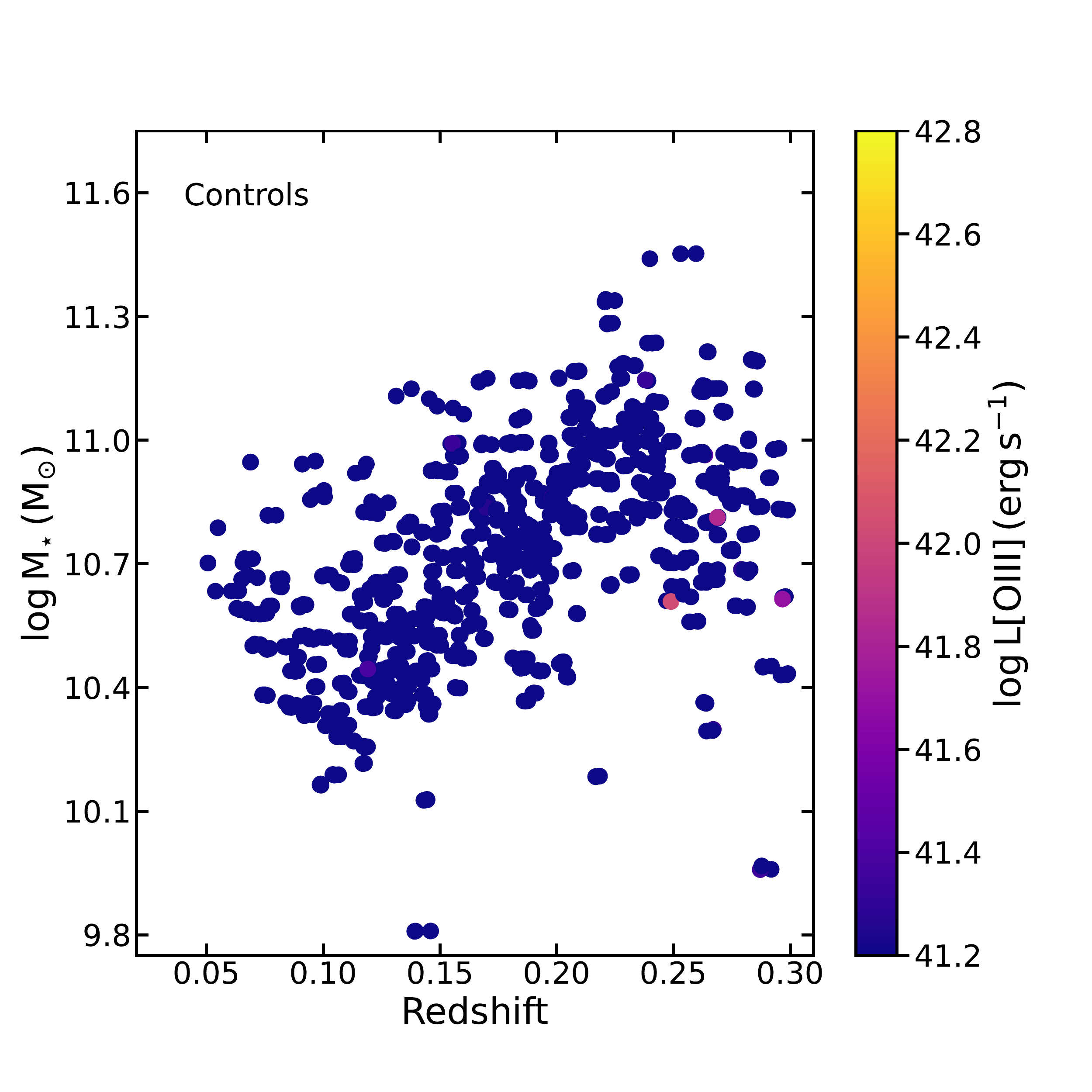}
    \caption{Stellar mass as a function of the redshift for the galaxies in our samples: QSOs (left panel) and control galaxies (right panel). Each point corresponds to a galaxy, and it is colour coded by the $\rm [O \, III] \, \lambda 5007$ luminosity of the galaxy. The control galaxies were selected to match the redshift and stellar mass of the QSOs, so we see a good agreement of both samples in these parameters. For a better comparison, the range of the colour bar is the same for both samples, therefore, all control galaxies with $\rm log \, L [OIII] < 41.2$ erg s$^{-1}$ are shown with the colour corresponding to $\rm log \, L [OIII] = 41.2$ erg s$^{-1}$ (blue). }
    \label{fig:mass-vs-z}
\end{figure*}

The QSO sample came from the catalogue of type 2 QSOs of \cite{2008AJ....136.2373R}. The use of type 2 QSOs for this study is motivated by the fact that, in order to investigate the galaxy morphology in SDSS-III images, due to its broad (FWHM $\sim $1.5 arcsec) PSF, it is important to avoid the glaring contribution of the AGN continuum and Broad-Line Region (BLR) that strongly affect the images of the host galaxies in type 1 AGN.

The original sample in the catalogue consists of 887 QSO 2s selected from the SDSS-III spectroscopic survey \citep{2011AJ....142...72E} with redshifts $z < 0.83$.
We performed a visual inspection of the r-band images and observed that the QSO hosts were unresolved at redshift above $\sim$ 0.3; we therefore restrict our sample to $z<0.3$. 
The resulting sample thus consists of 436 type 2 QSOs with $0.03 < z < 0.3$. 

In order to define control objects for our sample of QSOs, a necessary step is to characterise our sample according to the stellar mass of the QSO hosts. However, no such estimates have been made for the objects in our sample, due to the low contribution of the stellar population continuum in the SDSS-III fibre spectra of these objects. In order to estimate the masses of the hosts in our sample, we divided the 436 galaxies into ten redshift bins, after correcting all the individual spectra for galactic extinction. For each bin, we used the redshifts to bring the spectra to the rest frame and produced a stacked spectrum by median-combining the SDSS-III spectra from all QSOs in that bin. Using the software \textsc{starlight} \citep{2005MNRAS.358..363C,2006MNRAS.370..721M} we then performed a stellar population synthesis with a base of templates \citep{2003MNRAS.344.1000B} comprising 25 ages (1 Myr – 18 Gyr) and six metallicities ($0.0001 < Z < 0.05$),
and obtained the corresponding stellar mass $\overline{M}_{\star}$ contained in these combined fibre spectra. 
This mass is then used as a proxy for the stellar mass contained in the spectroscopic fibre of all galaxies that share the same redshift bin; even if this is not true for individual objects, the masses obtained are correct in a statistical sense. Finally, we convert the fibre stellar mass to total stellar mass $M_{\star}$ for each galaxy using their SDSS fibre (fibermag) and model (cmodelmag) $r$-band magnitudes, i.e.

\begin{equation}\label{eq: mass}
M_\star=\overline{M}_{\star}10^{0.4(\mbox{fibermag}-\mbox{cmodelmag})}   
\end{equation}

\noindent assuming a radially constant $r$-band $M/L$ ratio \citep{2021IAUS..359..339A}.

We have calculated the L$\rm [O \, III] \, \lambda 5007$ luminosity of the QSOs using the [OIII] flux ($f$) from table \texttt{galSpecLine} of SDSS-III DR12 \citep{2004MNRAS.351.1151B, 2004ApJ...613..898T} and its luminosity distance $D_{L}$. 

The left panel of Figure \ref{fig:mass-vs-z} shows the distribution of the 436 host galaxies of our QSO sample with respect to their stellar mass and redshifts and a range of colours indicating the L$\rm [O \, III]$ luminosity of the QSOs. 
The galaxies from our sample have thus stellar masses in the range $ 10^{9.8} M_{\odot} < {\rm M_{\star}} < 10^{11.7} M_{\odot}$ and luminosities in the range $\rm 10^{41.2}\, erg \, s^{-1} < L[OIII] < 10^{42.8}\,$erg$\,$s$^{-1}$.

\subsection{Control sample}

After the selection of the QSO sample, we also selected a control sample of non-active galaxies. With a control sample we can compare the results of our measurements regarding the environment and determine if there is a statistical difference between QSOs and matched galaxies without an AGN. 

The galaxies in the control sample were selected from the SDSS-III spectroscopic catalogue based on their stellar masses and redshifts only, that were matched to those of the QSOs. Objects that are considered as galaxies by the SDSS classifier and have estimated stellar masses listed in the \texttt{stellarMassPCAWiscBC03} table \citep{2003MNRAS.344.1000B, 2012MNRAS.421..314C} were potential candidates for the control sample.
We did not match the QSO and control galaxies according to their morphology. We have selected three control galaxies for each QSO, resulting in a sample of 1308 control galaxies, through the process described below.  

After separating the QSOs in 10 redshift bins and calculating their mean stellar masses as described in the previous section, for each bin,
we extracted from the SDSS-III spectroscopic sample a preliminary list of $\sim$ 10000 control candidates, in the same range of redshifts as the QSOs, and with stellar masses deviating at most 4$\sigma$ from the mean stellar mass of the QSOs in the bin.  

Using a least squares approach, we selected from this preliminary list the 3 galaxies that better matched each QSO from our sample. This was made via minimization of parameter $\alpha$ from Equation \ref{eq: match} below.  

\begin{equation}\label{eq: match}
\alpha = (z - z_0)^2 + (\log M - \log M_0)^2
\end{equation}

\noindent where $z$ and $M$ are the redshift and mass of the QSO host galaxy, and $z_0$ and $M_0$ are the redshift and mass of each galaxy in the preliminary list. 

In the right panel of Figure \ref{fig:mass-vs-z} we show the distribution of the control sample galaxies with respect to their stellar masses and redshifts. The colour represents again the $\rm [O \, III] \, \lambda 5007$ luminosity of the galaxy, as for the QSOs shown in the left panel.

\begin{figure}
    \centering
    \includegraphics[width=8cm]{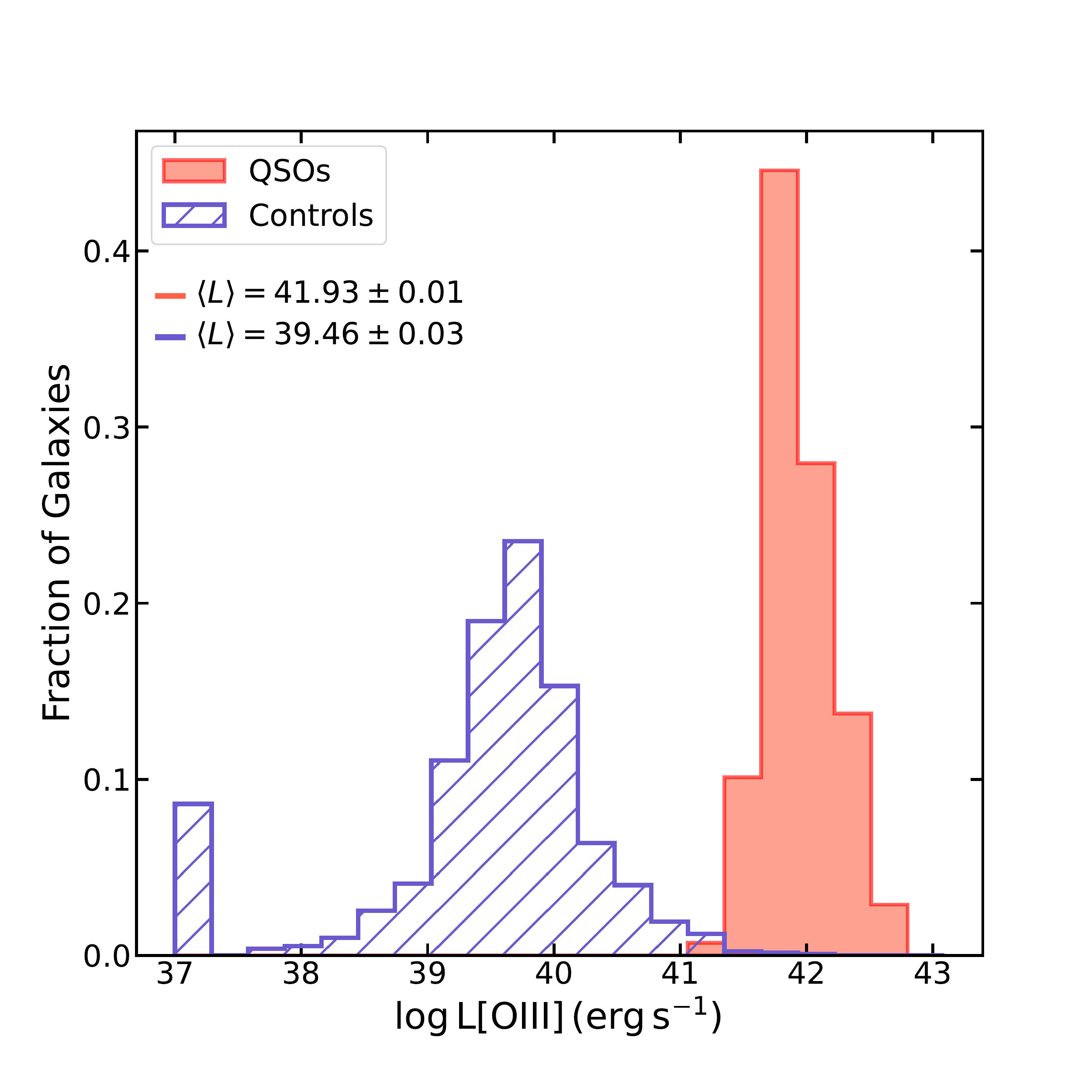}
    \caption{[O$\,$III] $\lambda 5007$ luminosity distribution for the QSO host galaxies (red) and control galaxies (blue) samples. The mean luminosity of each sample is displayed in the image $\langle L \rangle$. Galaxies with $\rm log \, L [OIII] < 37$ ergs/s are shown in the first bin of the plot. By selection, QSO host galaxies have higher emission line luminosities than control galaxies.}
    \label{fig:hist_lum}
\end{figure}

When selecting the control galaxies we 
did not use any cut in L$\rm [O \, III]$, which, as we can see in Figure \ref{fig:hist_lum}, is lower than those of the QSO hosts, with only a few with L$\rm [O \, III]$ above $10^{40}$ erg s$^{-1}$.

\section{Methodology}\label{sec: method}

In this section we will explain the methods used to calculate the following galaxy-environment properties for the two samples: the number of neighbours, tidal parameter and two asymmetry parameters. These properties were obtained using the SDSS images of the QSOs and controls.

\begin{figure*}
    \centering
    \includegraphics[width=17cm]{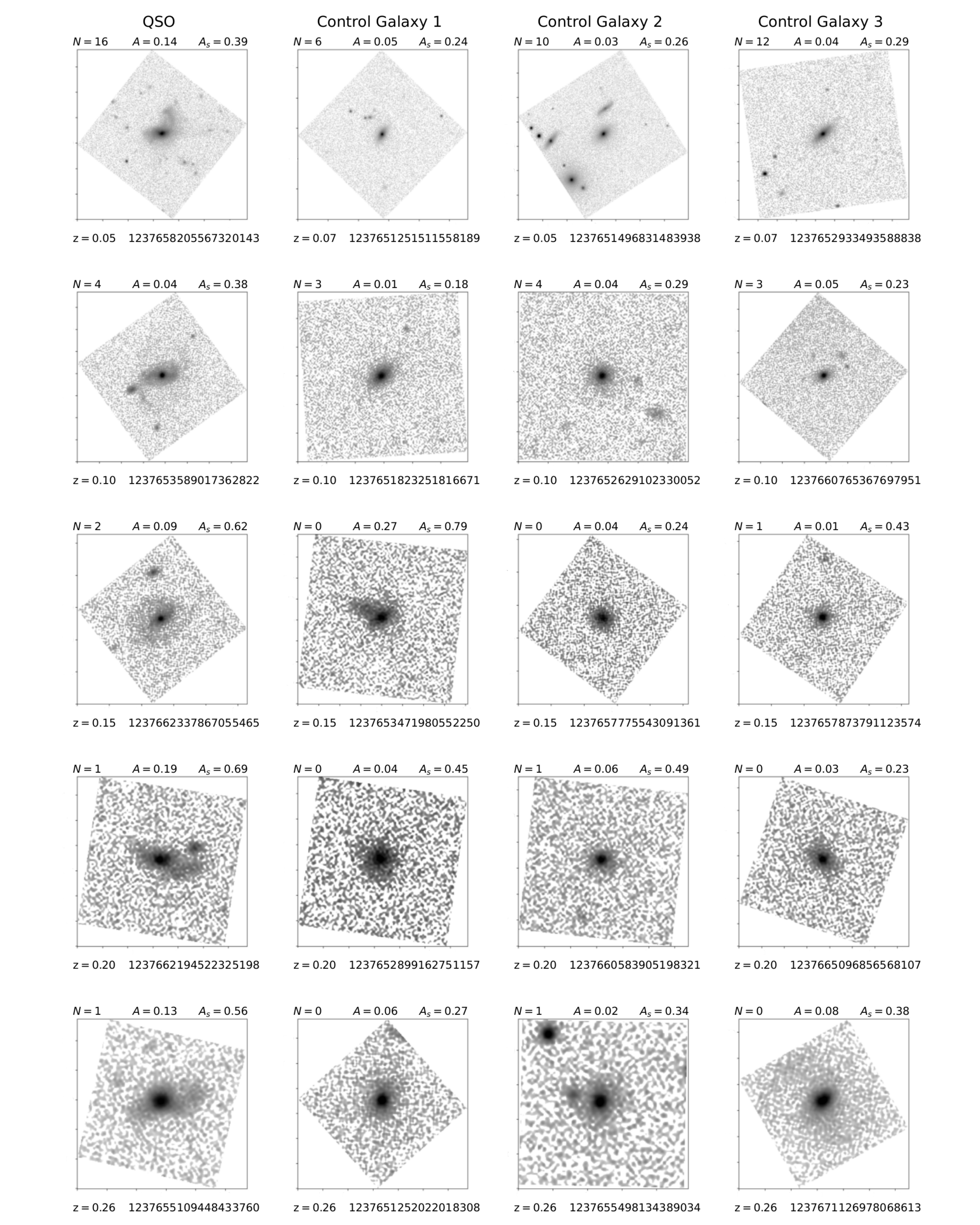}
    \caption{Examples of galaxies from our sample at different redshifts. The first column shows the QSOs and the three following columns show the corresponding control galaxies. Above each image we show the calculated value of the following parameters (see Section \ref{sec: method}): number of neighbours ($N$), asymmetry ($A$) and shape asymmetry ($A_s$). Bellow each image we show the redshift (z) of the respective galaxy and its identification (objid) in the SDSS catalogue. We have produced these images of the QSOs and their control galaxies with \textsc{SExtractor} using only the combined SDSS bands (ugriz) that are not contaminated by any emission line from the QSO. In all panels, the images (inner square) have a side of 100kpc.}
    \label{fig:fig_sample}
\end{figure*}

Since the galaxies in our sample cover a broad range of redshifts, the optical emission lines excited by the central engine will be detected in different wavelengths and will therefore leave a signature in different broadband filters for each QSO host. In order to reduce the contamination from the nuclear emission in the analysis of the (stellar) surface brightness distribution of the QSOs, we calculated the redshifted wavelength of the optical emission lines [O$\,$III] $\lambda 5007$ and H$\alpha$ for each host, and average-combined the SDSS-III images of each host in the filters whose passbands do not contain any of these lines. For the control galaxies, we combined the same images used to produce the combined image of their respective QSO host.
Figure \ref{fig:fig_sample} shows examples of the resulting images from 5 sets of QSO-controls from our sample, for different redshifts.

\subsection{Number of Neighbours}

We investigate whether QSO hosts are more/less likely to be physically associated to other galaxies -- i.e. to present a larger/lower frequency of neighbour objects -- than control galaxies. In order to quantify the frequency of neighbour objects around QSO hosts and control galaxies, it is necessary to identify all individual extended sources in the field of such objects. This step was performed on the combined images of each object with the code \textsc{SExtractor}\footnote{The parameters were set in such a way that the images could be visually detected, the parameters used were as follows: \texttt{DETECT\_THRESH} = 1.3, \texttt{DEBLEND\_NTHRESH} = 64, \texttt{DEBLEND\_MINCONT} = 0.001} \citep{1996A&AS..117..393B}. After detecting the secondary sources, we derived their integrated fluxes relative to that of the QSO host / control galaxy.

Galaxies in the foreground or background can contaminate the images and would be detected by the code as neighbours of the central galaxy. But we do not worry about this impacting our results because we use a control sample of galaxies, and the projection effect impacts both samples of galaxies, thus the comparison between them is still valid. 

To evaluate the environment at smaller distances, we also calculated the number of neighbours within 20$\,$kpc and 10$\,$kpc from the central galaxy, but the number of neighbours detected at these distances was too small that we could not do a statistical analysis of the results. Thus, we will not present them here.

\subsection{Tidal parameter}

The tidal strength parameter ($Q$) is a dimensionless number that estimates the total gravitational interaction that neighbours produce in a galaxy \citep{2015A&A...578A.110A}. The larger the value of $Q$, the less isolated from external influence is the central galaxy, and vice-versa.
We calculated the tidal strength parameter following equation \ref{eq:tidal},   

\begin{equation}\label{eq:tidal}
    Q \equiv \log \left[ \sum_{i} \frac{M_i}{M_c} \left( \frac{D_c}{d_i} \right) ^3 \right]
\end{equation}

\noindent where $M_c$ is the mass of the central galaxy, $D_c$ is the diameter of the central galaxy, $M_i$ is the mass of the neighbour galaxy and $d_i$ the projected distance from the neighbour galaxy to the central galaxy.

The mass of the central galaxy $M_c$ was already obtained through the stellar population synthesis using \textsc{starlight}, and to obtain the mass of the neighbour galaxy $M_i$ we used equation \ref{eq:mass}, since we have the fluxes ($f_c$ and $f_i$) from \textsc{SExtractor}.

\begin{equation}\label{eq:mass}
    \frac{M_i}{M_c} = \frac{f_i}{f_c}
\end{equation}

For the diameter of the central galaxy, we used $D_c = 2 \alpha R_{90}$, where $R_{90}$ is the Petrosian radius containing 90\% of the total flux of the galaxy in the SDSS $r$-band image and $\alpha = 1.43$ is scale factor used to recover D$_{25}$ from \citet{2013A&A...560A...9A}. 

We only take into  account the effect of neighbours that are closer than 50\,kpc from the central galaxy, as the $Q$ value becomes negligible beyond this radius. When there are no neighbour inside this radius, the tidal strength parameter is flagged as "NULL". 

\subsection{Asymmetry}

The presence of close companions around a galaxy can impact its stellar orbits and result in a disturbed morphology. One way to access such disturbances is to determine the degree of rotational asymmetry of the galaxy -- i.e. how much the galaxy image deviates from its galaxy-centred, 180$^\circ$-rotated counterpart. In this work we calculate two types of asymmetry. The first, that we call ``classic asymmetry'' or just ``asymmetry'', takes into account the brightness distribution of the source, in the sense that brighter regions contribute more to the resulting asymmetry value. The second type of asymmetry, called ``shape asymmetry'', considers only the asymmetry in the overall distribution of pixels occupied by the galaxy, no matter their relative intensities. We explain below how we calculate these two types of asymmetry.

\subsubsection{Classic asymmetry}

To calculate the classic asymmetry ($A$), we follow the standard definition of rotational asymmetry \citep{1996MNRAS.279L..47A, 2003ApJS..147....1C, 2008AJ....136.2115H}, as given by Equation \ref{eq:asymm}

\begin{equation}\label{eq:asymm}
    A = \frac{\sum |I_0 - I_\theta|}{\sum |I_0|} - A_f
\end{equation}

\noindent where $I_0$ is the flux from an individual pixel of the galaxy, and $I_\theta$ is the flux from a pixel in the same position, but of an image rotated by 180$^\circ$ around a centre point (selected to minimise the values of A) in the galaxy. $A_f$ is a correction due to the sky noise. The asymmetry $A$ has been shown to be sensitive both to disturbed morphologies and to small-scale features typically found in late-type spirals \citep[e.g.][]{1996MNRAS.279L..47A}. 

The calculation of $A$ was performed on the combined images of each object and did not take into account pixels closer to the galaxy centre than the average FWHM of the PSF, being therefore mostly insensitive to the contribution of eventual residual gas emission in the nuclear, unresolved region. Also, we only used pixels contained in a circular region within two Petrosian radii ($r_p$) around the centre of each galaxy. The internal one is the PSF radius and the external one is 2$r_p$, in a compromise between a good sampling of the peripheral regions and a reduction of noise due to sky-dominated pixels.

\subsubsection{Shape asymmetry}

Additionally to the classic asymmetry, we calculated the shape asymmetry ($A_s$), defined by \cite{2016MNRAS.456.3032P}.
To compute $A_s$, we use the same expression used for $A$ (Equation \ref{eq:asymm}), but now we use binary detection masks instead of the galaxies' images, meaning that all pixels that belong to the galaxy have the same "flux" value, and the pixels that don't belong to the galaxy are not taken into account.
Thus, $A_s$ is mostly sensitive to faint structures at the border of a galaxy like those produced by tidal interactions, being therefore an interesting tool to reveal an ongoing or past merger event.
Also, we do not restrict the calculation of $A_s$ at any given radial extent, relying instead on the full segmentation masks produced by \textsc{SExtractor} for each galaxy.

\section{Results}\label{sec:result}
\subsection{Number of neighbours}
\begin{figure}
    \centering
    \includegraphics[width=8cm]{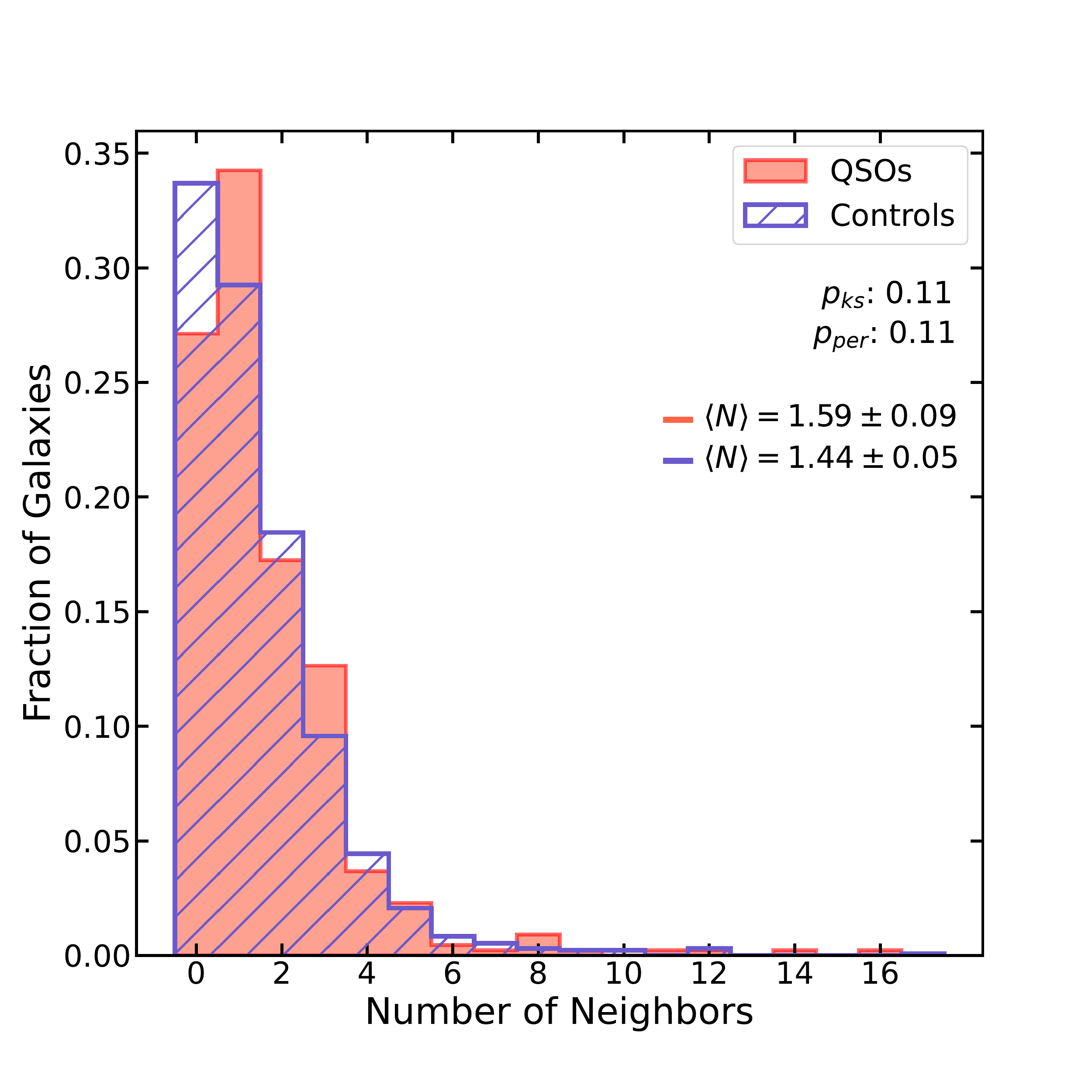}
    \caption{Distribution of the number of neighbours for the QSO (red) and control (blue) samples. We also include the p-values from two statistical tests: \textit{ks} and and permutation test. The mean $\langle N \rangle$ and the standard error of the mean of each distribution are also given.}
    \label{fig:nnei_hist}
\end{figure}

In Figure \ref{fig:nnei_hist} we show the histogram for the number of neighbours within 50\,kpc from the QSOs and control galaxies of our samples. Approximately 33\% of the control galaxies and $\sim 27\%$ of QSO host galaxies do not have any neighbour. In Figure \ref{fig:nnei_hist} we also give the mean $\langle N \rangle$ number of neighbour galaxies and the standard error of the mean for each distribution. The p-values for the Kolmogorov-Smirnov (\textit{ks}) statistics and permutation (\textit{permut} or \textit{per}) tests (both bigger than 0.05) -- also given in the Figure, show that there is no difference between the two samples in terms of the number of neighbours, even though the mean value is slightly higher for the QSO sample than for the control galaxies.  

\begin{figure*}
    \centering
    \includegraphics[width=17cm]{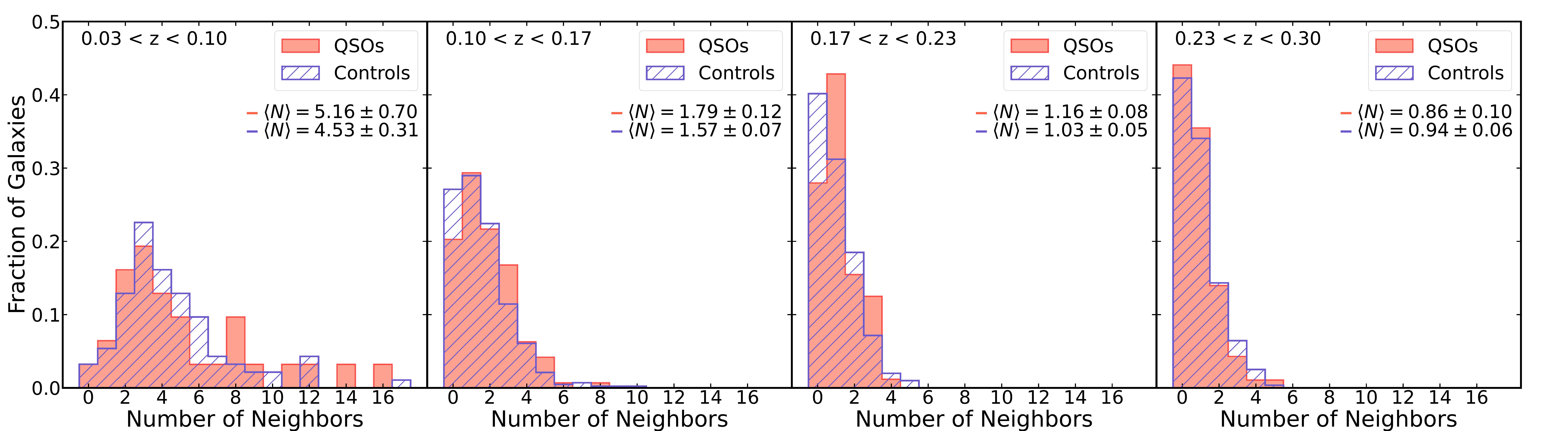}
    \caption{Histograms showing the number of neighbours in each sample separated in four ranges of redshift, shown in the top of each panel. The mean number of neighbours $\langle N \rangle$ and standard error of each sample is also given in the figure. We see that the mean number of neighbours is bigger at lower redshifts, as expected since the detection of galaxies becomes more difficult at higher redshifts.}
    \label{fig:nnei_hist_zbins}
\end{figure*}

The number of neighbours seen in the images should show a negative correlation with the redshifts of the galaxies since it gets harder to detect faint neighbours for farther away sources. To investigate this correlation we divided our sample into four sub-samples according to the redshift of the galaxies. Figure \ref{fig:nnei_hist_zbins} shows the distribution of the number of neighbours -- QSO host galaxies in red and control galaxies in blue, for each redshift interval - given at the top of each panel. As expected, the number of neighbours reaches higher values in the first redshift bin and lower values as the redshift increases. Table \ref{tab:nnei_z} gives additional information about each panel, such as the number of galaxies, mean number of neighbours and statistical test results for each redshift interval. But again, the statistical tests do not indicate a difference between QSO hosts and control galaxies in any of the redshift bins.  

\begin{figure*}
    \centering
    \includegraphics[width=17cm]{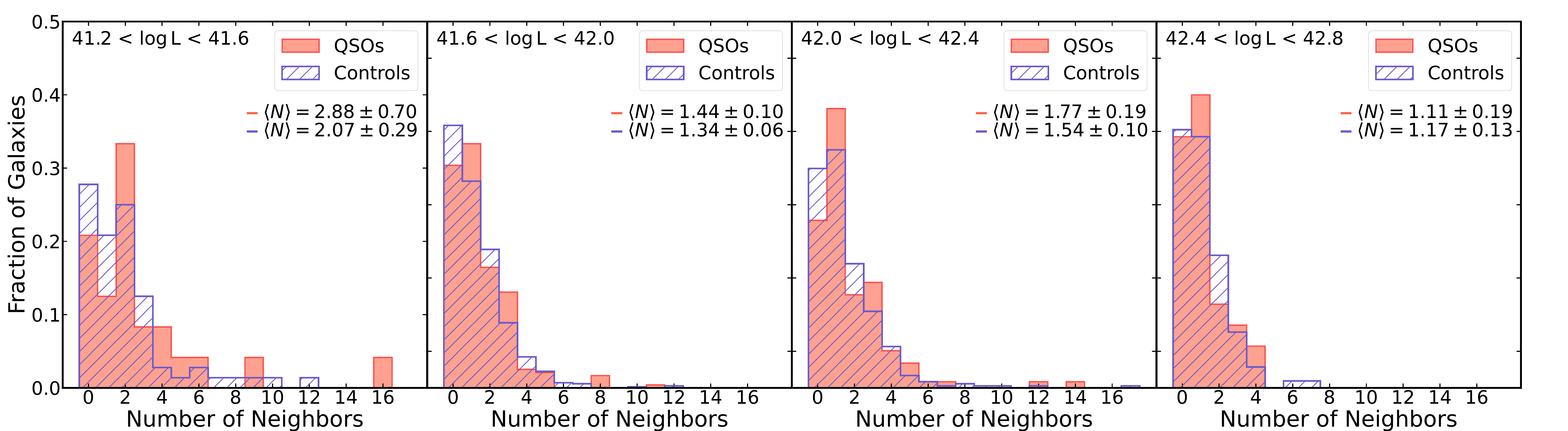}
    \caption{Histograms showing the number of neighbours in each sample in four ranges of L[O$\,$III] (in erg s$^{-1}$), from lower luminosity (left) to higher luminosity (right). The luminosity range is given at the top of each panel, as well as the mean number of neighbours $\langle N \rangle$ and standard error. We do not see a statistical difference between the two samples in any of the luminosity ranges. 
    }
    \label{fig:nnei_hist_lbins}
\end{figure*}

In order to investigate if there is a correlation between the [O$\,$III] emission line luminosity L[O$\,$III] of the QSOs and the number of neighbours, we also divided our sample in four bins of luminosity. Figure \ref{fig:nnei_hist_lbins} shows the histograms for the number of neighbours of the QSOs in the four bins of L[O$\,$III] and its respective control galaxies, with L[O$\,$III] increasing from left to right. Table \ref{tab:nnei_l} lists the number of galaxies, mean number of neighbours and results from statistical tests for each luminosity interval, showing that the less luminous sample presents the highest number of neighbours. But again there is no statistical difference between the number of neighbours of the QSO hosts and control galaxies in terms of L[O$\,$III] in any of the bins.

\begin{table*}
\centering
\caption{Table with information about the number of neighbours in the subsamples of Figure \ref{fig:nnei_hist_zbins}. This table summarises the redshift range, number of galaxies in each sample ($\rm n_q$ for QSOs and $\rm n_c$ for control galaxies), and mean number of neighbours (and standard error of the mean) in the QSOs and control samples. The results of statistical tests comparing the two samples are also shown, with p-values indicating the significance of the difference in neighbour statistics.}
\begin{tabular}{ l c c c c c c}
\hline
\multirow{2}{*}{\textbf{Redshift interval}} & \multicolumn{2}{c}{\textbf{Subsample}}& \textbf{QSOs} & \textbf{Controls} & \multicolumn{2}{c}{\textbf{p-values}}\\
& $\rm n_q$ & $\rm n_c$ & $\langle N \rangle$ & $\langle N \rangle$ & $\rm p_{ks}$ & $\rm p_{per}$\\\hline
0.03 $< z <$ 0.10 & 31 & 93 & 5.16 $\pm$ 0.69 & 4.53 $\pm$ 0.31 & 0.81 & 0.34\\[0.1cm]
0.10 $< z <$ 0.17 & 143& 428 & 1.79 $\pm$ 0.12 & 1.57 $\pm$ 0.07 & 0.60 & 0.13\\[0.1cm]
0.17 $< z <$ 0.23 & 168& 503 & 1.16 $\pm$ 0.08 & 1.03 $\pm$ 0.05 & 0.04 & 0.16\\[0.1cm]
0.23 $< z <$ 0.30 & 93& 279 & 0.86 $\pm$ 0.10 & 0.94 $\pm$ 0.06 & 1.00 & 0.49\\
\hline
\label{tab:nnei_z}
\end{tabular}
\end{table*}

\begin{table*}
\centering
\caption{Table with information about the number of neighbours in the subsamples of Figure \ref{fig:nnei_hist_lbins}. This table summarises the luminosity range, number of galaxies in each sample ($\rm n_q$ for QSOs and $\rm n_c$ for control galaxies), and mean number of neighbours (and standard error of the mean) in the QSOs and control samples. The results of statistical tests comparing the two samples are also shown, with p-values indicating the significance of the difference in neighbour statistics.}

\begin{tabular}{ l c c c c c c}
\hline
\multirow{2}{*}{\textbf{Luminosity interval}} & \multicolumn{2}{c}{\textbf{Subsample}}& \textbf{QSOs} & \textbf{Controls} & \multicolumn{2}{c}{\textbf{p-values}}\\
& $ \rm n_q$ & $\rm n_c$ & $\langle N \rangle$ & $\langle N \rangle$ & $\rm p_{ks}$ & $\rm p_{per}$\\\hline
41.21 $< \rm log \, L [OIII] <$ 41.59 & 24& 72 & 2.88 $\pm$ 0.70 & 2.07 $\pm$ 0.29 & 0.77 & 0.22\\[0.1cm]
41.59 $< \rm log \, L [OIII] <$ 41.98 & 237& 709 & 1.44 $\pm$ 0.10 & 1.34 $\pm$ 0.06 & 0.64 & 0.36\\[0.1cm]
41.98 $< \rm log \, L [OIII] <$ 42.37 & 118& 354 & 1.77 $\pm$ 0.19 & 1.54 $\pm$ 0.10 & 0.76 & 0.24\\[0.1cm]
42.37 $< \rm log \, L [OIII] <$ 42.76 & 35& 105 & 1.11 $\pm$ 0.19 & 1.17 $\pm$ 0.13 & 1.00 & 0.76\\
\hline
\label{tab:nnei_l}
\end{tabular}
\end{table*}

\begin{figure}
    \centering
    \includegraphics[width=8cm]{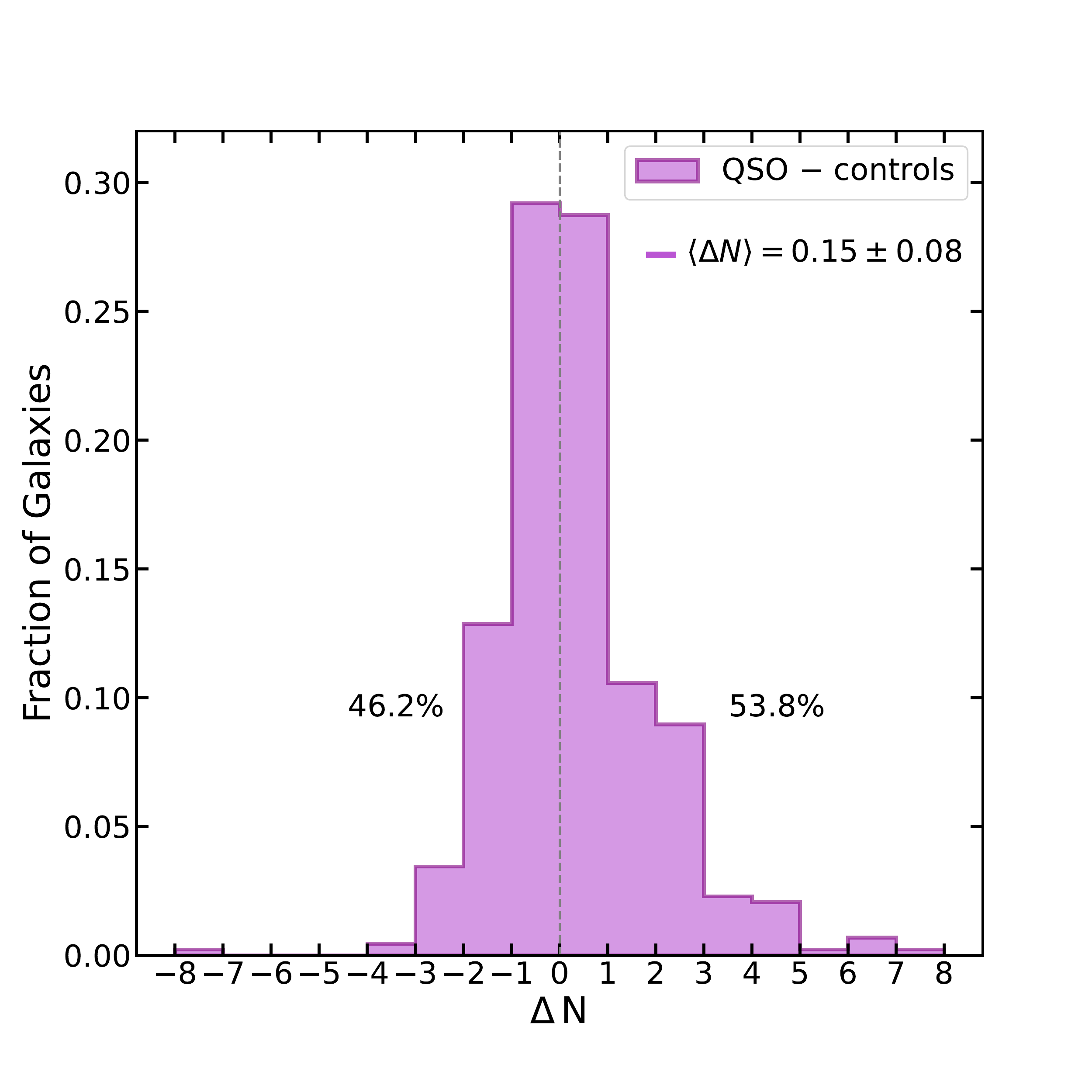}
    \caption{Histogram of $\Delta N$ (difference between the number of neighbours of a QSO and the mean number of neighbours of its three control galaxies) for all the galaxies in our sample. We include the mean value $\langle \Delta N \rangle$ and the standard error of the mean, as well as the percentage of QSOs showing higher number of neighbours than the controls (positive $\Delta N$) and lower number of neighbours than the control galaxies (negative $\Delta N$).}
    \label{fig:dif_nnei_hist}
\end{figure}

As we have paired each QSO with three control sample galaxies, we can also compare the results for each QSO with those for its respective control galaxies. We have thus calculated the difference $\Delta N$ between the number of neighbours of the QSO host galaxy and the mean number of neighbours of its three control galaxies. The histogram of $\Delta N$ is shown in Figure \ref{fig:dif_nnei_hist}, that has a mean $\langle \Delta N \rangle = 0.15 \pm 0.08$. This value, larger than zero, indicates a slightly larger number of neighbours in QSO hosts than in their control galaxies. Figure \ref{fig:dif_nnei_hist} also reveals that $\Delta N$ is larger than zero in 53.8\% of the QSO sample.

We also evaluated the distance to the closest companion for the galaxies in both samples, but we found no difference between QSOs and controls.

\subsection{Tidal parameter}

\begin{figure}
    \centering
    \includegraphics[width=8cm]{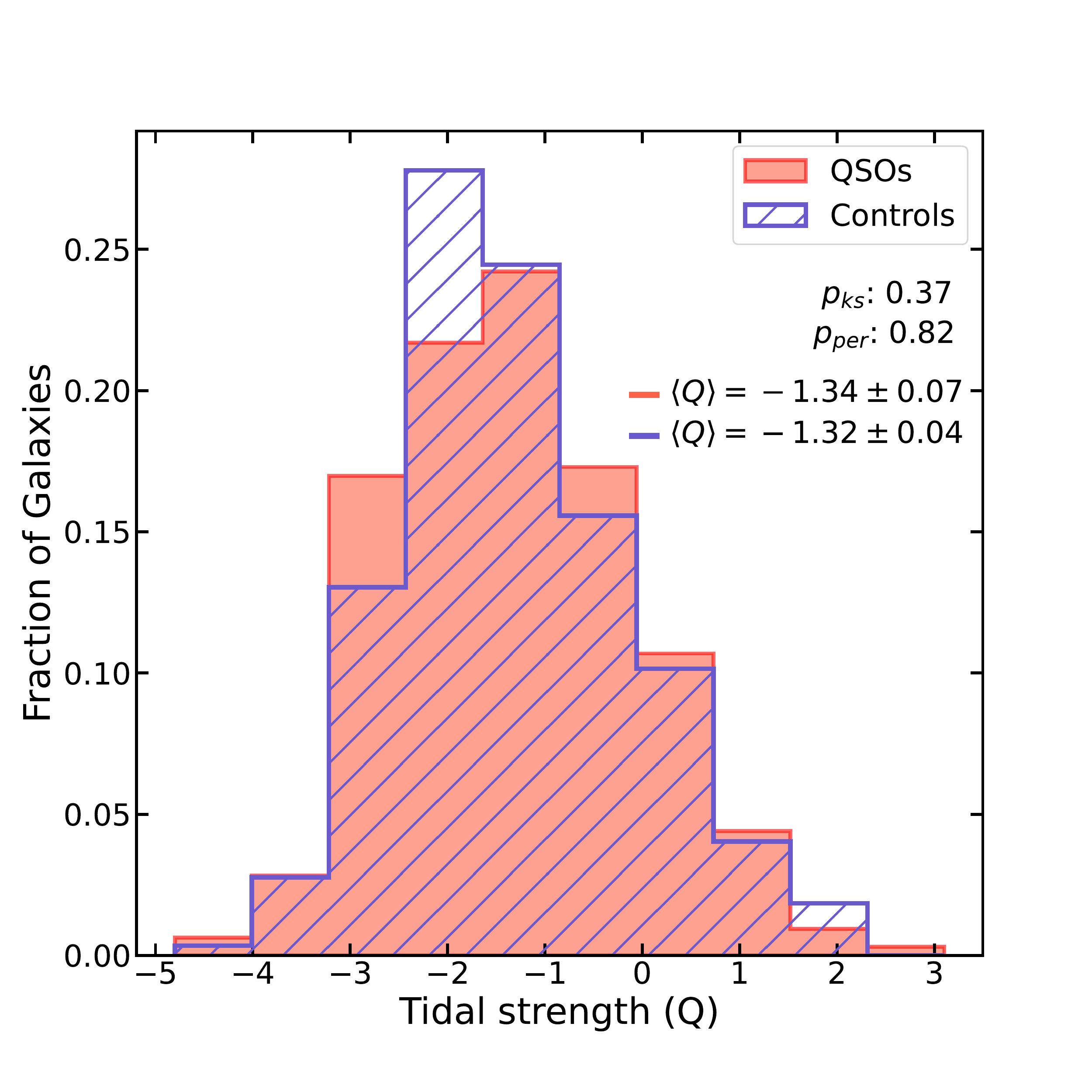}
    \caption{Tidal strength (Q) histogram for the QSO hosts sample (red) and the control sample (blue). The p-values of the tests \textit{ks} and \textit{permut} indicate that the two samples are not statistically different. We also show the mean $\langle Q \rangle$ and the standard error of the mean of each distribution. }
    \label{fig:tidal_hist}
\end{figure}

Figure \ref{fig:tidal_hist} shows the histogram of the tidal strength parameter (Q) for the galaxies in the QSO sample (red) and control sample (blue). In the figure we also list the result of the statistical tests and the mean value of Q for each sample $\langle Q \rangle$. We see no statistical difference between the two samples and we also don't see a difference in the mean.

\subsection{Asymmetry}

We calculated the asymmetry for all 436 QSO hosts and 1308 control galaxies, using the method described in section \ref{sec: method}. But, to ensure the reliability of the data, we removed from the sample galaxies with classical asymmetry higher than 0.3 or lower than -0.2, and/or galaxies with shape asymmetry higher than 1 (these galaxies correspond to $\sim$\,4\% of the galaxies from each sample). A visual inspection of the corresponding images showed that these extreme values are obtained when the image is too noisy, or when there is a star too close to the galaxy, and thus it is not possible to calculate accurately the asymmetry. The final sample thus consists of 417 QSO hosts, and 1254 control galaxies.

We now present the results for the classical and shape asymmetries of our sample.

\subsubsection{Classic asymmetry}

\begin{figure}
    \centering
    \includegraphics[width=8cm]{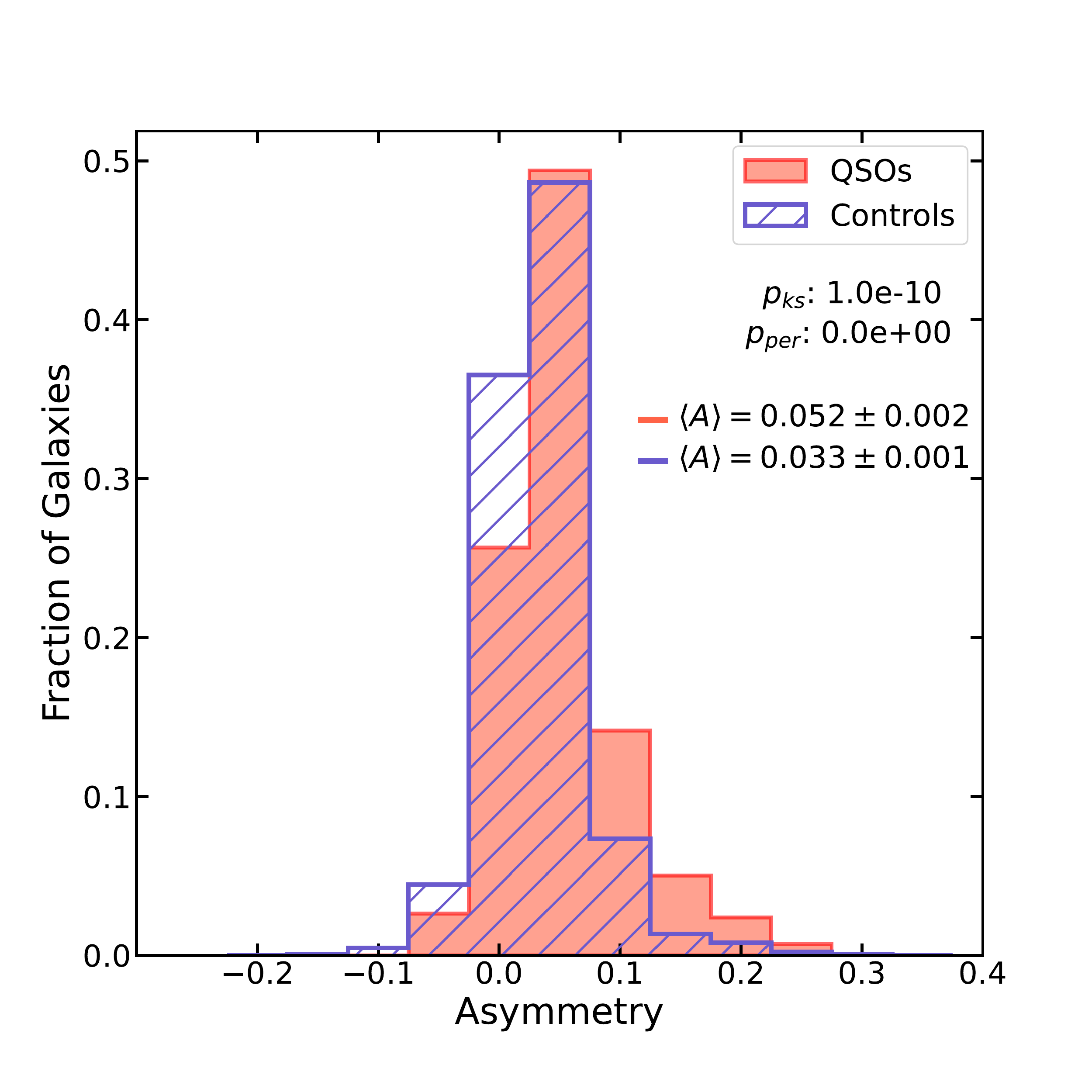}
    \caption{Distribution of the asymmetry for the QSO hosts (red) and control (blue) samples. The p-values of the \textit{ks} and \textit{permut} tests indicate that the two samples are statistically different. We also list the mean $\langle A \rangle$ and the standard error of the mean of each distribution. The mean asymmetry of the QSO hosts is almost two times bigger than that of the control galaxies.}
    \label{fig:asymm_hist}
\end{figure}

Figure \ref{fig:asymm_hist} shows the distribution of the galaxies according to their asymmetry values, QSO host galaxies in red and control galaxies in blue. The p-value of the statistical tests are also given in the figure showing very low values, demonstrating that the QSO and control samples are statistically different. The mean asymmetry of each distribution is also listed in the figure. The QSO host galaxies are on average more asymmetrical than the control galaxies.

\begin{figure*}
    \centering
    \includegraphics[width=16cm]{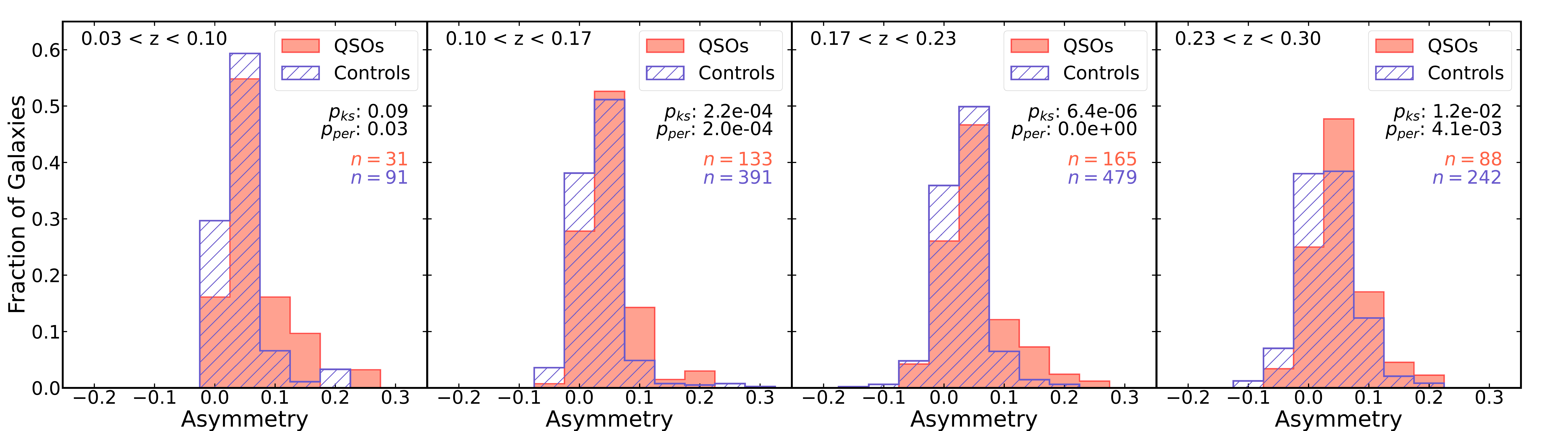}
    \caption{Histograms showing the asymmetry values for each sample in four ranges of redshift. The redshift for each plot is listed at the top of each panel, as well as the number of objects from each sample in the given range of redshift. The asymmetry of the QSO hosts is larger in all ranges of redshift, in agreement with the mean values listed in Table \ref{tab:asymm_z}.
    The statistical tests indicate a difference between the two samples in all ranges of redshift, except in the first one ($\rm 0.03 < z < 0.10$) where the p-value of the \textit{ks} test is above 0.05, but this might be due to the low number of objects in this redshift range.}
    \label{fig:asymm_hist_zbins}
\end{figure*}

Figure \ref{fig:asymm_hist_zbins} shows histograms of the asymmetry separated in four bins of redshift. 
QSO host galaxies are represented in red and control galaxies in blue. The p-values of the statistical tests are given in the figures, and Table \ref{tab:asymm_z} summarises additional information of each subsample. Except for the first bin, there is a clear difference between the QSO sample and control sample, as evidenced by the p-values of the tests, i.e $\rm p_{ks}$ and $\rm p_{per}$ that are smaller than 0.05, with the QSO sample showing a higher mean value for the asymmetry in all redshift bins. For the first redshift interval, $0.03<z<0.10$, the p-value of the \textit{ks} test is larger than 0.05, but given the small number of galaxies in this bin, the test can not give a confident result. None the less, the mean asymmetry of the QSO sample is higher than that of the control sample for this bin as well. 

\begin{figure*}
    \centering
    \includegraphics[width=16cm]{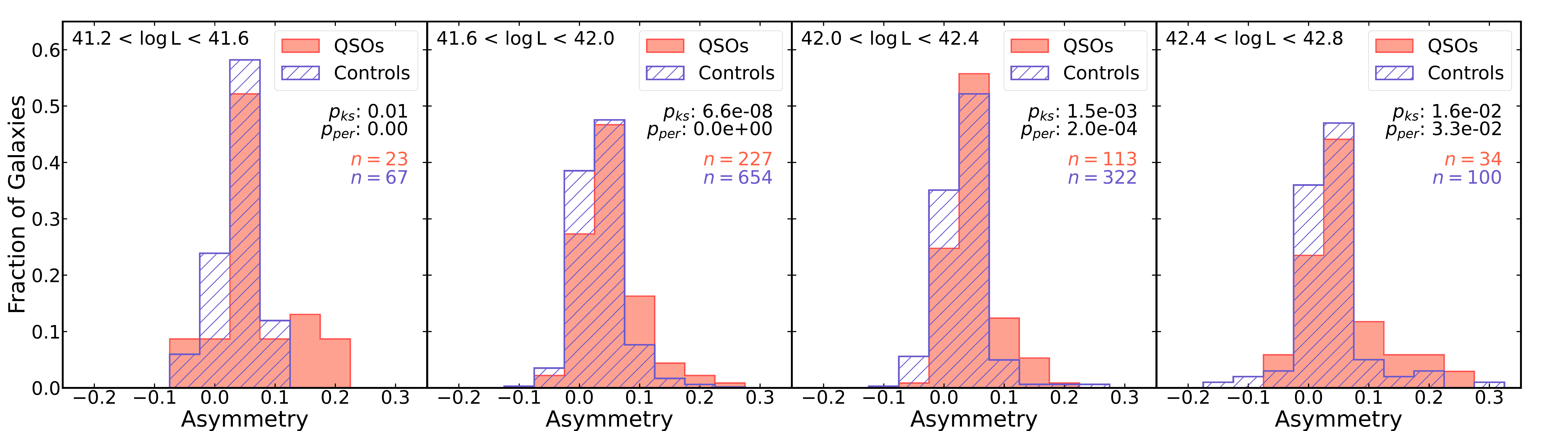}
    \caption{Histograms showing the asymmetry in each sample in four ranges of log\,L[O$\,$III] (in erg s$^{-1}$), from lowest luminosity (left) to highest luminosity (right). The luminosity ranges are shown at the top of each panel, and the number of objects from each sample in the given range of luminosity are also listed, together with the results of the statistical tests. The asymmetry of the QSO hosts is larger in all plots, as given by the mean values listed in Table \ref{tab:asymm_l}. The statistical tests indicate a difference between the two samples in all ranges of luminosity.}
    \label{fig:asymm_hist_lbins}
\end{figure*}

We also divided the sample in four bins of L[O$\,$III]. Histograms of the asymmetry of the galaxies separated in these luminosity bins are shown in Figure \ref{fig:asymm_hist_lbins}. The luminosity interval (in erg s$^{-1}$) and the number of galaxies in each interval is listed at the top of each panel. Again, we applied statistical tests to compare both samples and calculated the mean and standard error of the mean for each distribution, presented in Table \ref{tab:asymm_l}, as well as the number of galaxies in each subsampple and the results for the statistical testes. All tests indicate a clear difference between QSO hosts and control galaxies, and the mean asymmetry of the QSO hosts is higher than the mean asymmetry of the control galaxies in all luminosity bins. 

\begin{table*}
\centering
\caption{Table with information about the asymmetry in the subsamples of Figure \ref{fig:asymm_hist_zbins}. This table summarises the redshift range, number of galaxies in each sample ($\rm n_q$ for QSOs and $\rm n_c$ for control galaxies), and mean asymmetry (and standard error of the mean) in the QSOs and control samples. The results of statistical tests comparing the two samples are also shown, with p-values indicating the significance of the difference in asymmetry statistics.}
\begin{tabular}{ l c c c c c c}
\hline
\multirow{2}{*}{\textbf{Redshift interval}} & \multicolumn{2}{c}{\textbf{Subsample}}& \textbf{QSOs} & \textbf{Controls} & \multicolumn{2}{c}{\textbf{p-values}}\\
& $ \rm n_q$ & $\rm n_c$ & $\langle A \rangle$ & $\langle A \rangle$ & $\rm p_{ks}$ & $\rm p_{per}$\\\hline
0.03 $< z <$ 0.10 & 31& 91 & 0.063 $\pm$ 0.009 & 0.043 $\pm$ 0.004 & 0.09 & 0.03\\[0.1cm]
0.10 $< z <$ 0.17 & 133& 391 & 0.049 $\pm$ 0.004 & 0.033 $\pm$ 0.002 & 0.00 & 0.00\\[0.1cm]
0.17 $< z <$ 0.23 & 165& 479 & 0.053 $\pm$ 0.004 & 0.032 $\pm$ 0.002 & 0.00 & 0.00\\[0.1cm]
0.23 $< z <$ 0.30 & 88& 242 & 0.049 $\pm$ 0.005 & 0.032 $\pm$ 0.003 & 0.01 & 0.00\\
\hline
\label{tab:asymm_z}
\end{tabular}
\end{table*}

\begin{table*}
\centering
\caption{Table with information about the asymmetry in the subsamples of Figure \ref{fig:asymm_hist_lbins}. This table summarises the luminosity range, number of galaxies in each sample ($\rm n_q$ for QSOs and $\rm n_c$ for control galaxies), and mean asymmetry (and standard error of the mean) in the QSOs and control samples. The results of statistical tests comparing the two samples are also shown, with p-values indicating the significance of the difference in asymmetry statistics.}
\begin{tabular}{ l c c c c c c}
\hline
\textbf{Luminosity interval} & \multicolumn{2}{c}{\textbf{Subsample}}& \textbf{QSOs} & \textbf{Controls} & \multicolumn{2}{c}{\textbf{p-values}}\\
$\rm erg \, s^{-1}$& $ \rm n_q$ & $\rm n_c$ & $\langle A \rangle$ & $\langle A \rangle$ & $\rm p_{ks}$ & $\rm p_{per}$\\\hline
41.21 $< \rm log \, L [OIII] <$ 41.59 & 23& 67 & 0.068 $\pm$ 0.013 & 0.035 $\pm$ 0.004 & 0.01 & 0.00\\[0.1cm]
41.59 $< \rm log \, L [OIII] <$ 41.98 & 227& 654 & 0.052 $\pm$ 0.003 & 0.033 $\pm$ 0.002 & 0.00 & 0.00\\[0.1cm]
41.98 $< \rm log \, L [OIII] <$ 42.37 & 113& 322 & 0.049 $\pm$ 0.004 & 0.033 $\pm$ 0.002 & 0.00 & 0.00\\[0.1cm]
42.37 $< \rm log \, L [OIII] <$ 42.76 & 34& 100 & 0.061 $\pm$ 0.011 & 0.035 $\pm$ 0.006 & 0.02 & 0.03\\
\hline
\label{tab:asymm_l}
\end{tabular}
\end{table*}

\begin{figure}
    \centering
    \includegraphics[width=8cm]{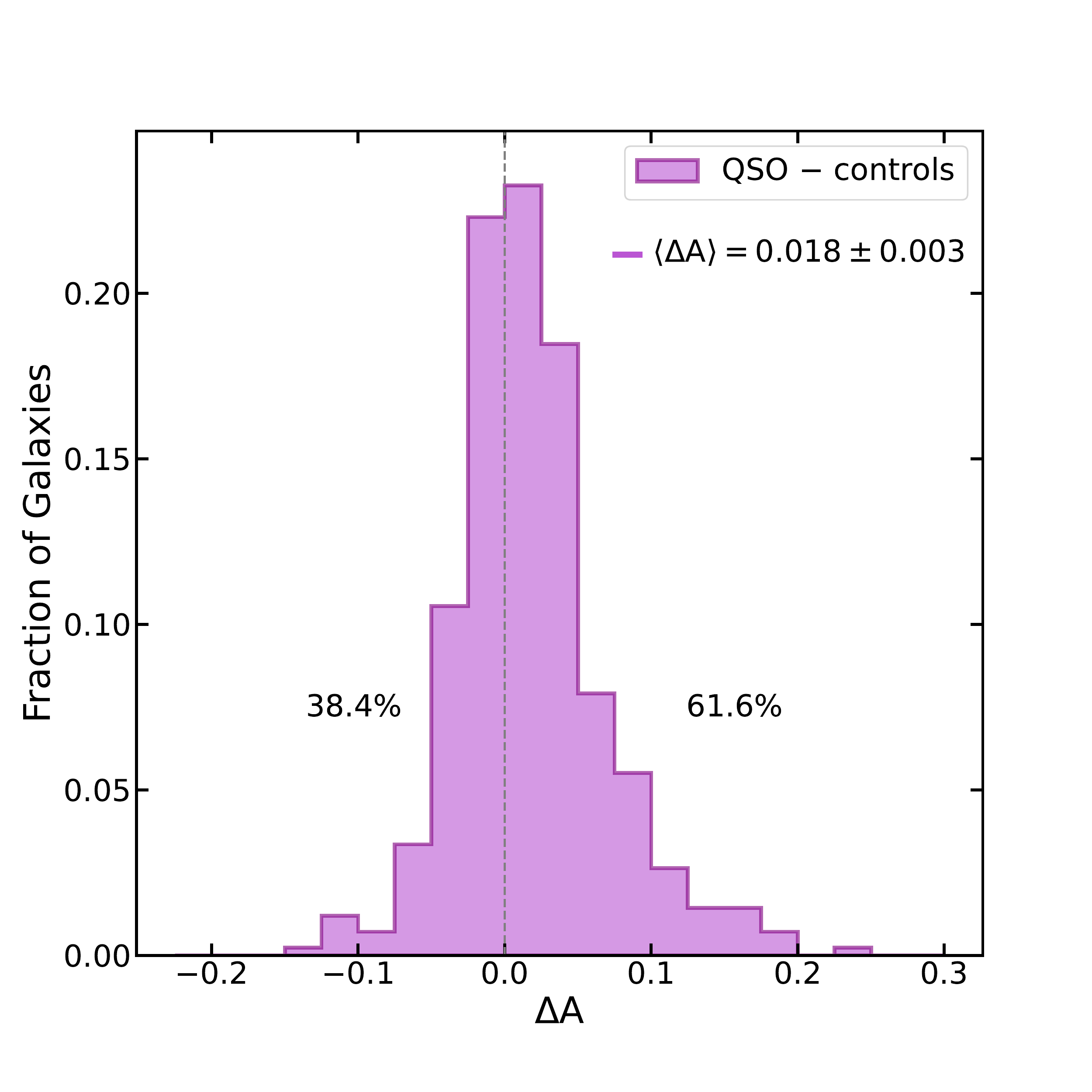}
    \caption{Histogram of $\Delta A$ (difference between the asymmetry of a QSO and the mean asymmetry of its three control galaxies) for all galaxies of our sample. The mean $\langle \Delta A \rangle$ and the standard error of the mean of the distribution are also listed. The mean value is above zero, showing that on average, the QSO host galaxies are more asymmetrical than their control galaxies.}
    \label{fig:asymm_hist_dif}
\end{figure}

To directly compare each QSO host galaxy with their three control galaxies, we calculated the parameter $\Delta A$, the difference between the asymmetry of the QSO host and the mean asymmetry of its three control galaxies. Figure \ref{fig:asymm_hist_dif} shows the histogram for $\Delta A$. From the mean value ($\langle \Delta A \rangle = 0.018 \pm 0.003$) and the shape of the histogram we can see that most of the QSO hosts are more asymmetrical than their control galaxies.
We counted the number of sets (QSO-controls) for which the asymmetry of the QSO is larger than that of its control galaxies and found that this occurs in $61.6\%$ (257 from 417) of the sample.

\begin{figure*}
    \centering
    \includegraphics[width=8cm]{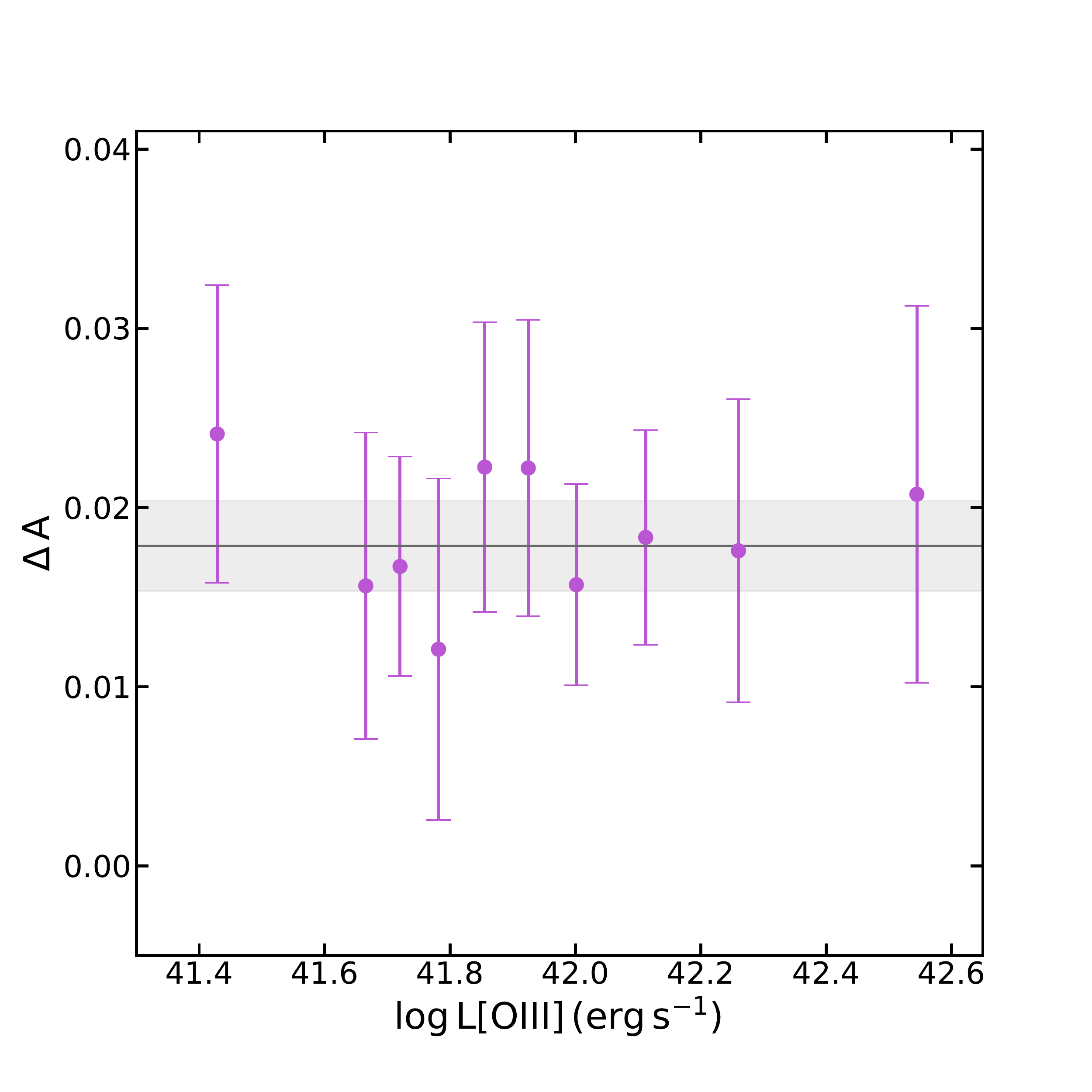}
    \includegraphics[width=8cm]{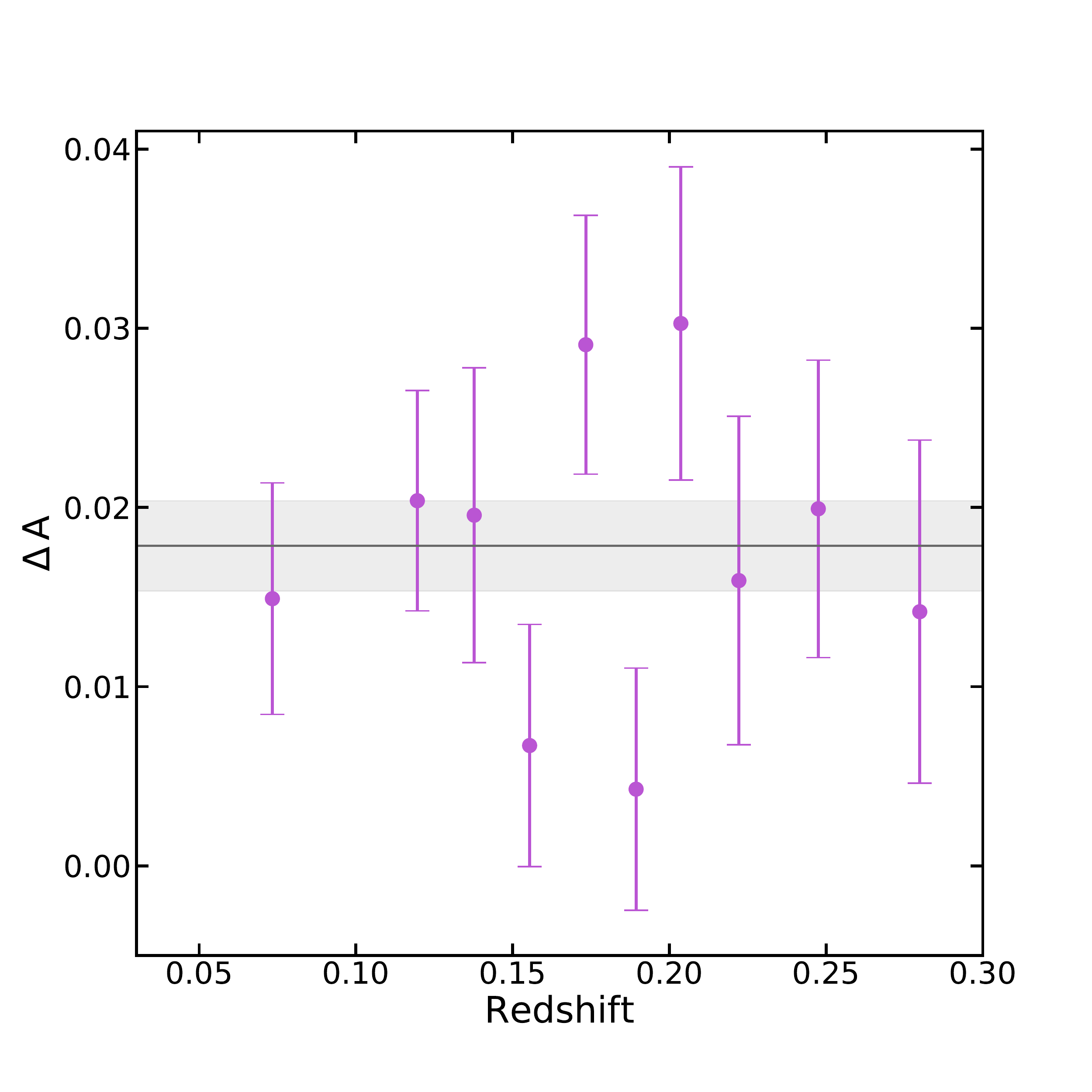}
    \caption{The purple points show the mean values of $\Delta A$ in 10 bins of log\,L[O$\,$III] (left panel) and redshift (right panel). 
    The error bars correspond to the standard error of the mean. The horizontal line and shaded area corresponds to the mean and standard error of the mean for the complete sample ($\langle \Delta A \rangle = 0.018 \pm 0.003$). All points are above zero, and there is no apparent trend with respect to L[O$\,$III] or redshift.}
    \label{fig:delta_asymm}
\end{figure*}

In Figure \ref{fig:delta_asymm} we investigate the relation between $\Delta A$ and L[O$\,$III] (left panel), and the redshift (right panel). We divided the sample in 10 bins with an equal number of objects in each bin. The points in the figure show the mean value of $\Delta A$ for each bin. The error bars correspond to the standard error of the mean $\Delta A$. The horizontal line in the plots represents the mean $\Delta A$ of the complete sample ($ \langle \Delta A \rangle = 0.018 \pm 0.003$), and the shaded area is the standard error of the mean. We note that all points are above zero, which means that in all bins (both in luminosity and redshift) the QSO host galaxies are on average more asymmetrical than their control galaxies. We do not see any trends in $\Delta A$ as a function of the QSO luminosity or redshift.

\subsubsection{Shape asymmetry}

\begin{figure}
    \centering
    \includegraphics[width=8cm]{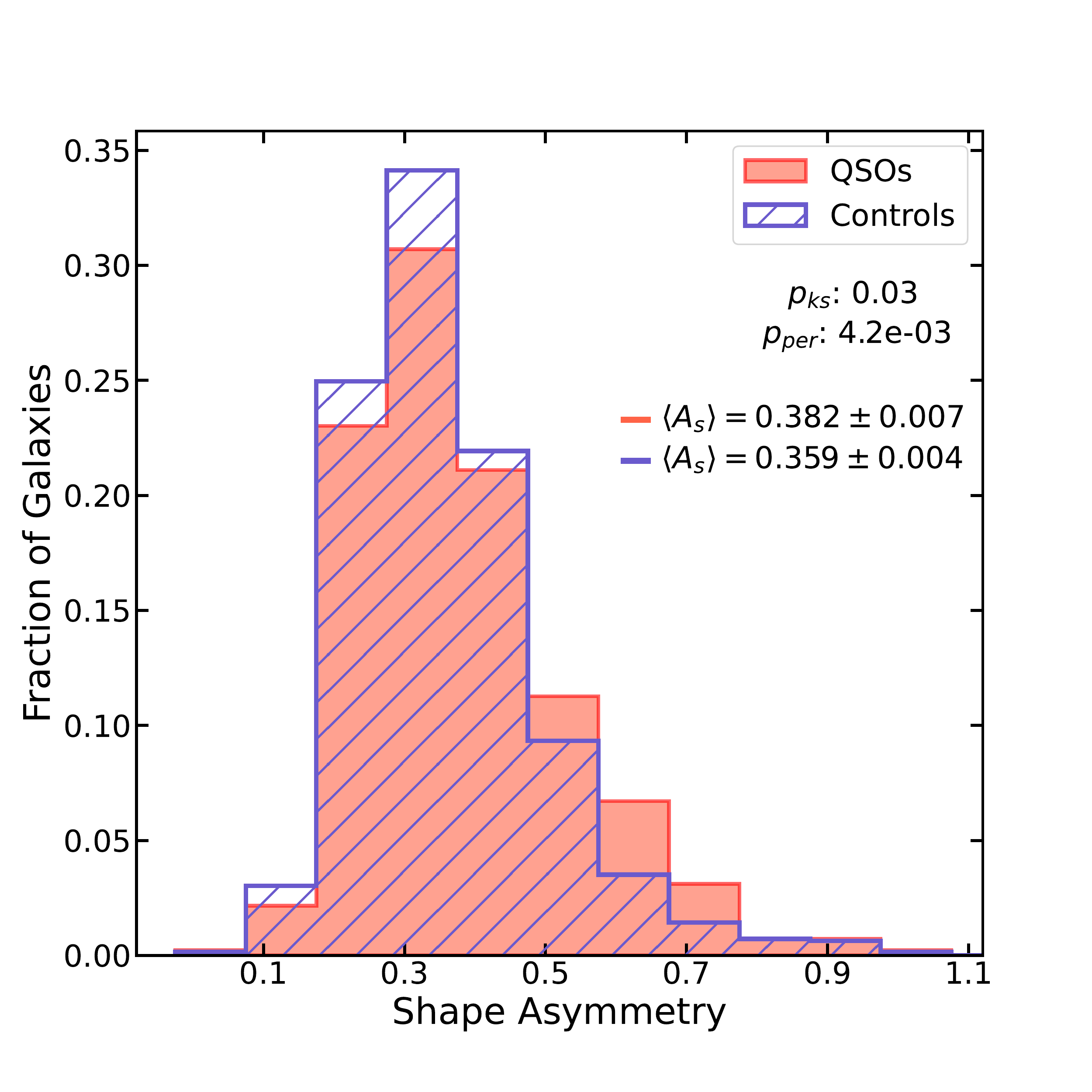}
    \caption{Distribution of the shape asymmetry for the QSO hosts (red) and control sample (blue). The p-values of the tests \textit{ks} and \textit{permut} indicate that the two samples are statistically different. We also show the mean values $\langle A_s \rangle$ and the standard error of the mean for each distribution. The mean shape asymmetry of the QSO host galaxies is higher than that of the control galaxies.}
    \label{fig:shape_asymm_hist}
\end{figure}

Figure \ref{fig:shape_asymm_hist} shows the histogram for the shape asymmetry $A_s$ calculated for the QSO hosts (in red) and control galaxies (in blue). The p-values of the statistical tests are shown in the figure, together with the mean and standard error of the mean for each distribution. From the p-values it can be concluded that the two samples are statistically different, with the QSO host galaxies presenting a higher mean shape asymmetry than the control galaxies. 

We also divided the sample in bins of redshift and luminosity, as done for the classic asymmetry, but we did not see a statistical difference between the two samples, and we thus do not show here the corresponding histograms.

\begin{figure}
    \centering
    \includegraphics[width=8cm]{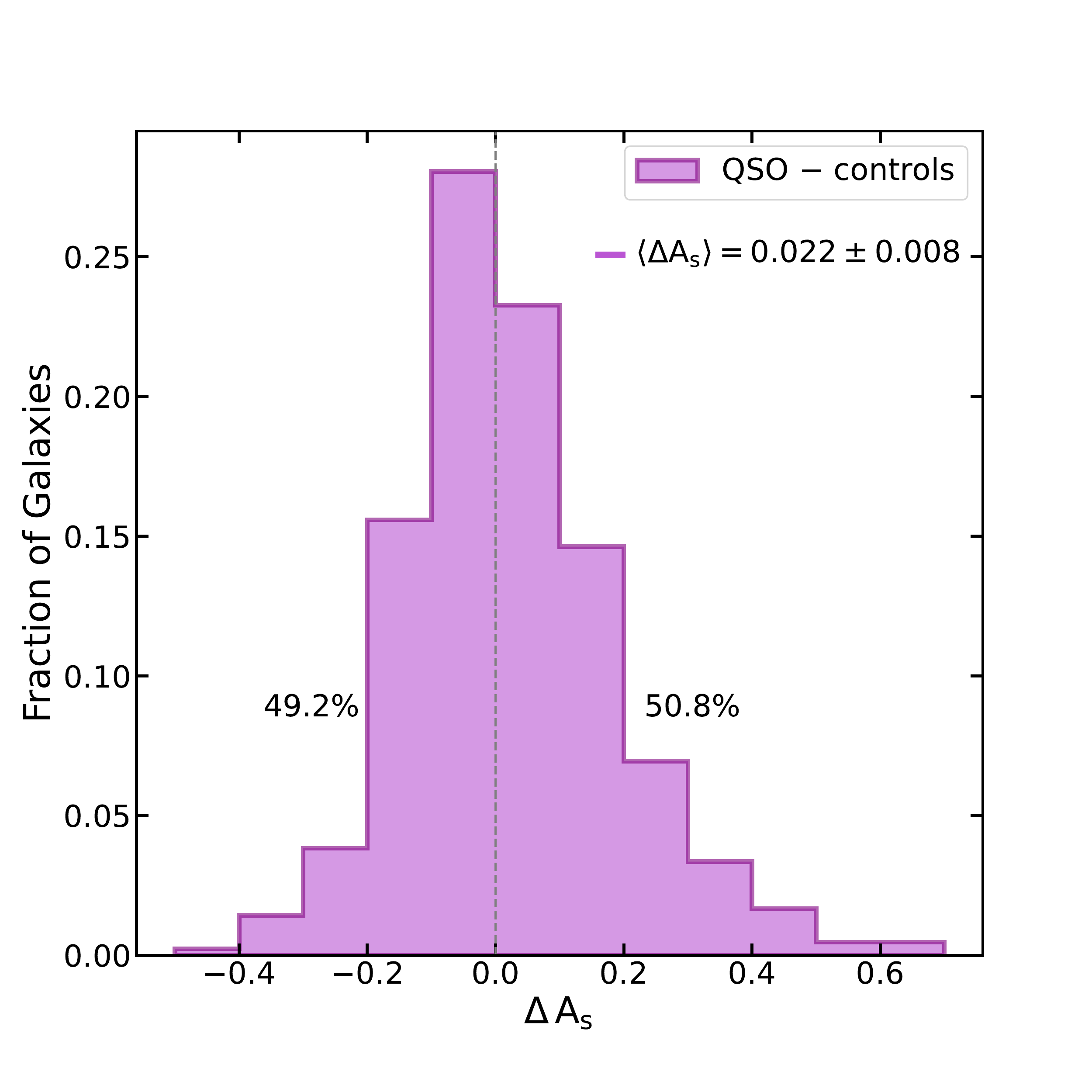}
    \caption{Histogram of $\Delta A_s$ (difference between the shape asymmetry of each QSO and the mean shape asymmetry of its three control galaxies) for all the galaxies in our sample. In the figure we also list the mean $\langle \Delta A_s \rangle$ and the standard error  of the mean of the distribution, together with the percentage of cases in which $\Delta A_s$ is higher and lower than zero.
    }
    \label{fig:shape_asymm_hist_dif}
\end{figure}

To directly compare each QSO host galaxy to its three control galaxies, we calculated the parameter $\Delta A_s$, the difference between the shape asymmetry of the QSO host and the mean shape asymmetry of its three control galaxies. Figure \ref{fig:shape_asymm_hist_dif} shows the histogram for this quantity. The mean value ($\langle \Delta A_s \rangle = 0.022 \pm 0.008$) indicates that most of the QSO hosts are, on average, more asymmetrical in shape than their control galaxies, but $\Delta A_s$ for each set of QSO - control galaxies is small. We calculated the number of sets (QSO-controls) with shape asymmetry of the QSO higher than the mean shape asymmetry of its control galaxies and found that in only $50.8\%$ (212 from 417) of them this is the case.

\section{Discussion}\label{sec: discuss}

We have investigated the environment of the QSOs via the comparison  of the following measured properties (using continuum images of the host galaxies and their environment) with those of a control sample: number of neighbours $N$, tidal parameter $Q$ and two types of asymmetry, the classical asymmetry $A$ and shape asymmetry $A_s$. We find a small excess in $N$ only when we compare each QSO with its three control galaxies and found no difference in $Q$. The main difference was found in the asymmetry parameters $A$ and $A_s$. The difference is more significant in the classical asymmetry $A$ -- that takes into account the light distribution in the galaxies' images, than in the shape asymmetry, that takes into account only the difference in the morphology of the images. We now discuss the meaning of these results.

\subsection{Lack of difference in number of neighbours and tidal parameter}

In our analysis, we investigated the number of neighbours around the host galaxies of QSO\,2's and control galaxies, as shown in Figures \ref{fig:nnei_hist} to \ref{fig:dif_nnei_hist}. To identify neighbours, we searched within a 50\,kpc radius around the galaxy using SDSS-III optical images. It is important to note that our ability to detect close neighbours is limited by the sensitivity of the survey; and as a result, faint galaxies may not have been detected. Additionally, since we observed a wide range of redshifts ($0.03<z<0.3$), it is expected that closer galaxies will have a higher number of detectable neighbours compared to more distant ones. Despite these limitations, we matched the QSO host galaxies with control galaxies based on their redshift and stellar mass, allowing us to compare the two samples in a consistent manner. Our analysis did not reveal any significant differences in the number of neighbours between the QSO host galaxies and control galaxies. 

Our results suggests that, within a radius of 50\,kpc, the presence of neighbouring galaxies may not significantly impact the central region of the host galaxy. However, our study does reveal a difference in asymmetry between the QSO and control samples -- as further discussed below, indicating that close enough neighbouring galaxies (within distances of the order of the galaxy radius) play a role in the activity of the SMBH. These findings are consistent with previous studies that have found no significant differences in the large-scale environments of AGN and control galaxies, but have found differences when examining smaller distances, typically on the order of 30\,kpc \citep{2016ApJ...832..111J,2018MNRAS.479.2308B}.

Figure \ref{fig:tidal_hist} shows that there is no significant difference in the tidal parameter between the QSOs and control galaxies in our sample. This is not surprising, as the tidal parameter depends on the number of neighbouring galaxies, as well as their distances and masses. Since we did not find a significant difference in the number of neighbours between the two samples (as discussed in the previous paragraph), it is expected that the tidal parameter would also be similar. Therefore, our results suggest that the presence of neighbouring galaxies at distances of $\sim$50\,kpc from the QSOs does not have a significant impact on the tidal forces experienced by the galaxy, and thus may not play a major role in the triggering of nuclear activity in these systems.

\subsection{Asymmetries}

\begin{figure*}
    \centering
    \includegraphics[width=14cm]{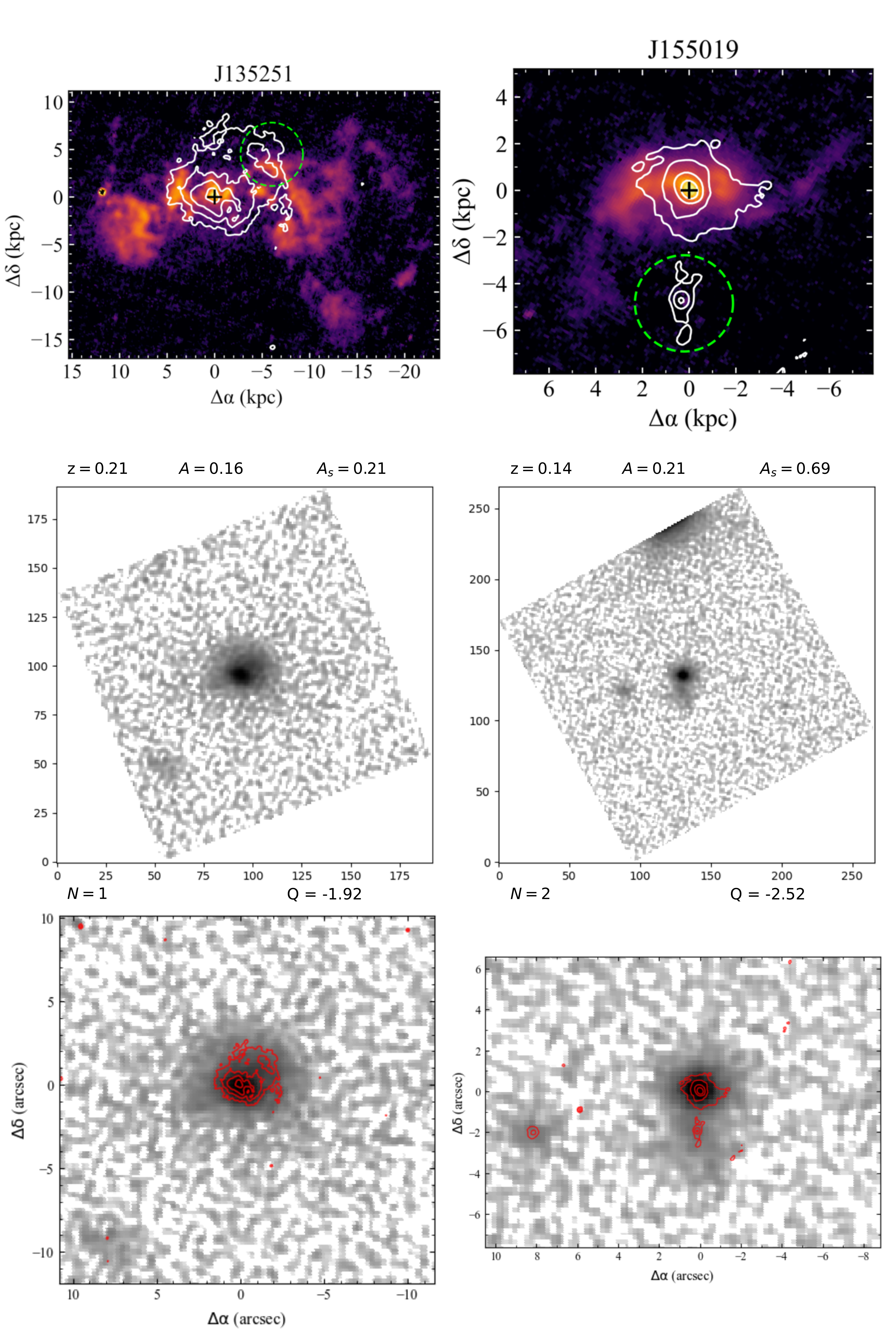}
    \caption{Comparison between HST QSO images and the SDSS images used in our work for two QSOs with high asymmetry values. The top HST images show the [O$,$III] flux distributions in orange to purple, with white contours from images of the stellar continuum both in logarithmic scale \citep{2021MNRAS.504.3890D}. The middle panel shows the SDSS images of the QSOs, which were made by combining the u, g, and z bands. The lower panels show the zoomed in SDSS images overlaid with the stellar contours shown in the HST images. The redshift (\textit{z}), asymmetry ($A$), shape asymmetry ($A_s$), number of neighbours (\textit{N}), and tidal parameter (\textit{Q}) values of each QSO are also provided.
    }
    \label{fig:qso_bruno}
\end{figure*}

Regarding the classical asymmetry, the QSO host galaxies are more asymmetrical than the control galaxies, with mean asymmetries for the QSOs hosts and control galaxies of, respectively,  $\langle A \rangle = 0.052 \pm 0.002$, and $\langle A \rangle = 0.033 \pm 0.001$. After dividing the sample into four bins of redshift and luminosity, the difference between the two samples persists (Figures \ref{fig:asymm_hist}, \ref{fig:asymm_hist_zbins} and \ref{fig:asymm_hist_lbins}). 

As for the shape asymmetry, the QSO host galaxies are just slightly more asymmetrical than the controls when considering the whole sample, with mean values for the QSO and controls of respectively, $\langle A_s \rangle = 0.382 \pm 0.007$ and $\langle A_s \rangle = 0.359 \pm 0.004$ (Figure \ref{fig:shape_asymm_hist}). But when we divide the sample into four bins of redshift and luminosity, the difference is not statistically significant for most bins.

When we compare the differences in asymmetry of each QSO and their controls, the QSOs are, on average, more asymmetrical for both asymmetries (Figures \ref{fig:asymm_hist_dif} and \ref{fig:shape_asymm_hist_dif}). 
But, while in the classical asymmetry, 61,4\% of the QSO are more asymmetrical than their controls, for the shape asymmetry, this percentage is only 50,8\%.

What is the meaning of the above results? Asymmetry in galaxies can be associated with interactions. The gravitational effect of a neighbour galaxy can disturb first the gas and then the stars in a galaxy creating, for example, tidal tails and even asymmetries in the inner regions. 

To better understand the meaning of the measured asymmetry, we have compared the continuum SDSS images we have used here with Hubble Space Telescope images available for two QSOs of our sample that have high values of asymmetry ($A$ = 0.16 and $A$ = 0.21). This comparison is shown in Figure \ref{fig:qso_bruno}, with HST images shown on the top, SDSS images in the middle, and the SDSS with superimposed contours of the HST continuum images in the bottom.
As we have used SDSS images in bands that avoid the contamination of strong lines, the comparison should be done with the continuum HST images.
The presence of neighbour galaxies in both QSOs can be seen in the HST continuum images, shown as contours in Fig.\ref{fig:qso_bruno}, with the neighbour galaxies highlighted by the green circles in the figure. For J135251, the neighbour galaxy is $\approx$ 5$\,$kpc north of the QSO host, and the two are already merging. In J155019, the neighbour galaxy is also $\approx$ 5$\,$kpc to the south the QSO host. In these two cases, Figure \ref{fig:qso_bruno} shows that the neighbour galaxies are so close to the QSO host that the SDSS images do not show them as separate galaxies. Thus in our analysis of these images, they are not identified as separate galaxies but, instead, as galaxies with high asymmetry: $A=0.16$ and $A_s=0.21$ and $A=0.21$ and $A_s=0.69$ for J135251 and J155019, respectively. 

The above comparison suggests that the excess in asymmetry found in QSOs are most probably due to the presence of close companions or mergers that cannot be identified in the SDSS images. 

And why is the excess in classical asymmetry greater than in the shape asymmetry? Considering that the classical asymmetry takes into account the flux distribution in the source, while this does not happen in the case of  the shape asymmetry, these results suggest that the triggering of the QSO only occurs when the disturbance has reached already the inner regions of the galaxy, that contributes most to the flux.

\subsubsection{Area of the galaxies}

In the two galaxies shown in Figure \ref{fig:qso_bruno} the companions are so close (5\,kpc) that they are already in the process of merging \footnote{While the redshift of the companions cannot be confirmed to be the same as that of the galaxy, the presence of clear asymmetries that seem to connect the two galaxies suggests that they are close to each other and even already in the process of merging.}.
In order to estimate this distance for the rest of the sample, that can probably be considered the typical distance at which a companion would trigger the nuclear activity, we have calculated the area covered by the stellar component of the galaxies that can also be obtained from our measurements using the program \textsc{SExtractor}.

\begin{figure}
    \centering
    \includegraphics[width=8cm]{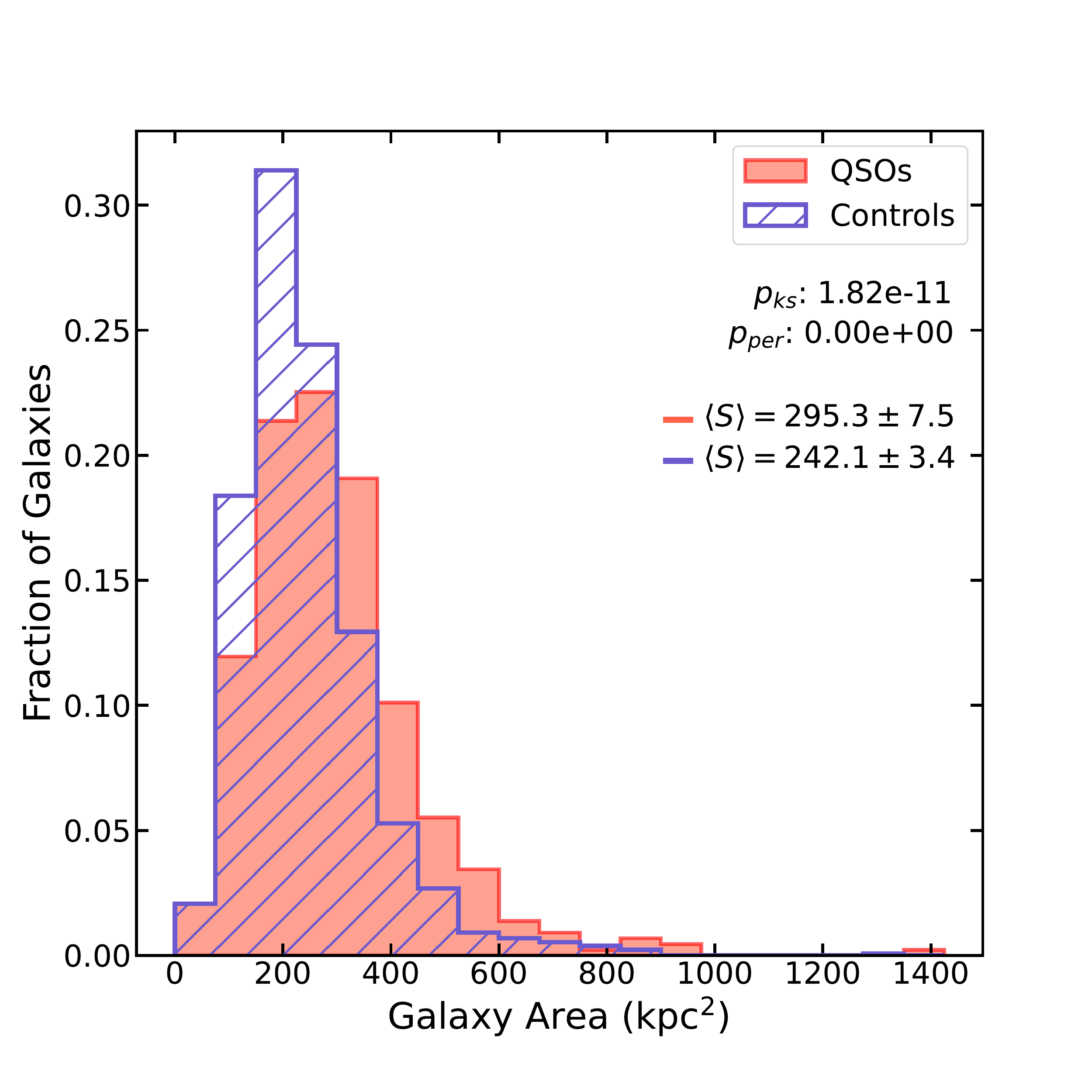}
    \caption{Histogram for the area of the QSOs (red) and control galaxies (blue), where the p-values and median values $\langle S \rangle$ for each sample are also given.}
    \label{fig:hist_gal_area}
\end{figure}

We have used the pixel count information given by \textsc{SExtractor} to calculate the area $S$ of the galaxies in both samples, given by

\begin{equation}
   S = n_{px} \left(\frac{0.398\,\pi}{3600\times 180}\,\mbox{rad\,pix$^{-1}$}\right) ^2DA(z)^2,
    \label{eq:area}
\end{equation} 

\noindent where $n_{px}$ is the number of pixels contained in the segmentation image of the galaxy, and \textit{DA(z)} is its angular size distance at redshift $z$.  In Figure \ref{fig:hist_gal_area}, the distribution of the values of these areas for the QSOs (red) and control galaxies (blue) shows that QSO host galaxies have, on average, a larger area when compared to control galaxies. The mean area $\langle S \rangle$ of the galaxies in each sample is also shown in the histogram, $\langle S \rangle = 295.3 \pm 7.5$\,kpc$^2$ and $\langle S \rangle = 242.1 \pm 3.4$\,kpc$^2$, for QSOs and controls, respectively. The p-values for the \textit{ks} and \textit{permut} tests show that the two samples are statistically different.

A larger area for the QSOs relative to the control supports the interpretation that the asymmetry is due to a close companion that is in a process of merger with the QSO galaxy.

\begin{figure}
    \centering
    \includegraphics[width=8cm]{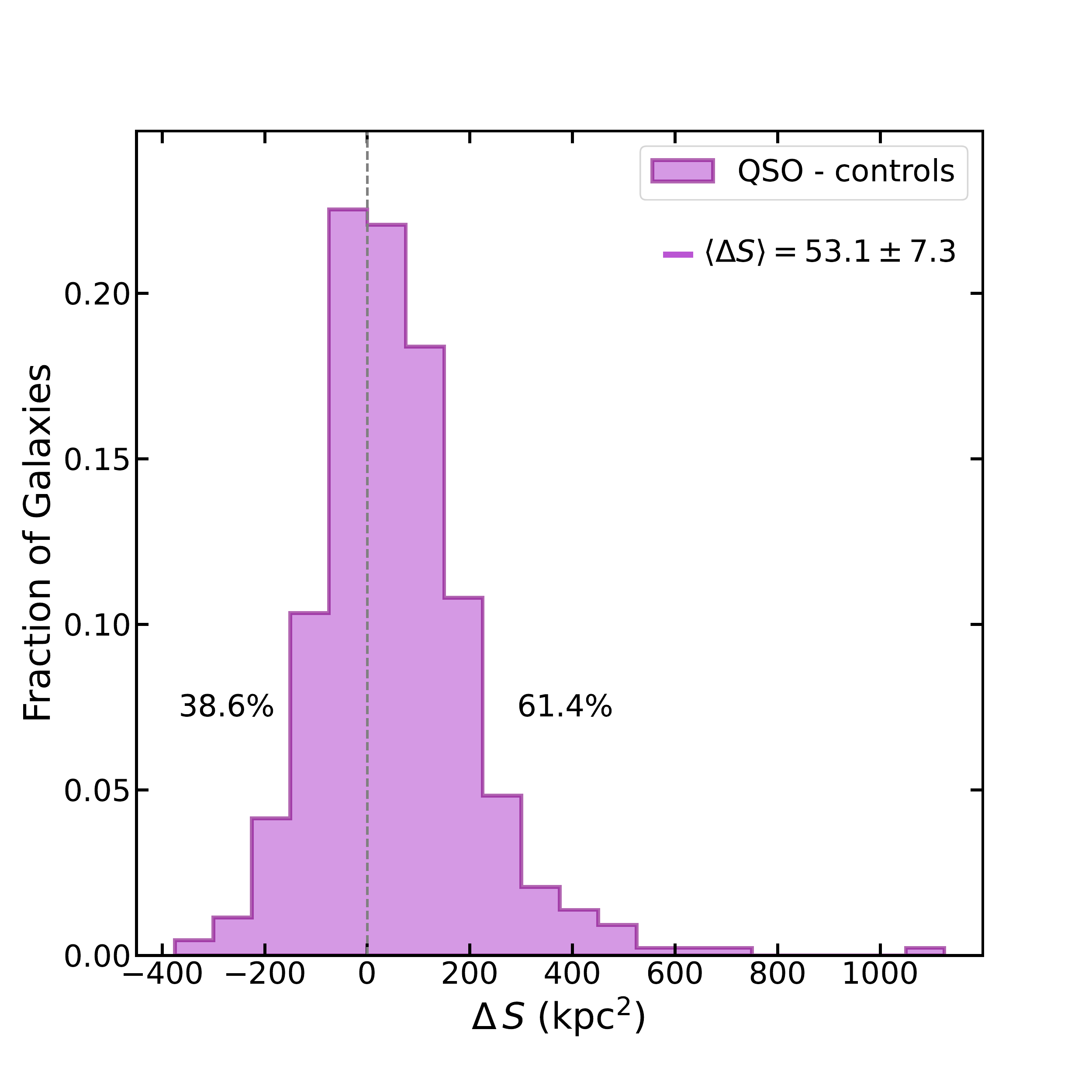}
    \caption{Histogram of $\Delta S$ (difference between the area of each QSO and the mean area of its three control galaxies) for all the galaxies of our sample. We also list the mean $\langle \Delta S \rangle$ and the standard error of the mean of the distribution. The mean is higher than zero, showing that on average, the QSO host galaxies have a larger area than its control galaxies.}
    \label{fig:hist_gal_area_dif}
\end{figure}

\begin{figure}
    \centering
    \includegraphics[width=8cm]{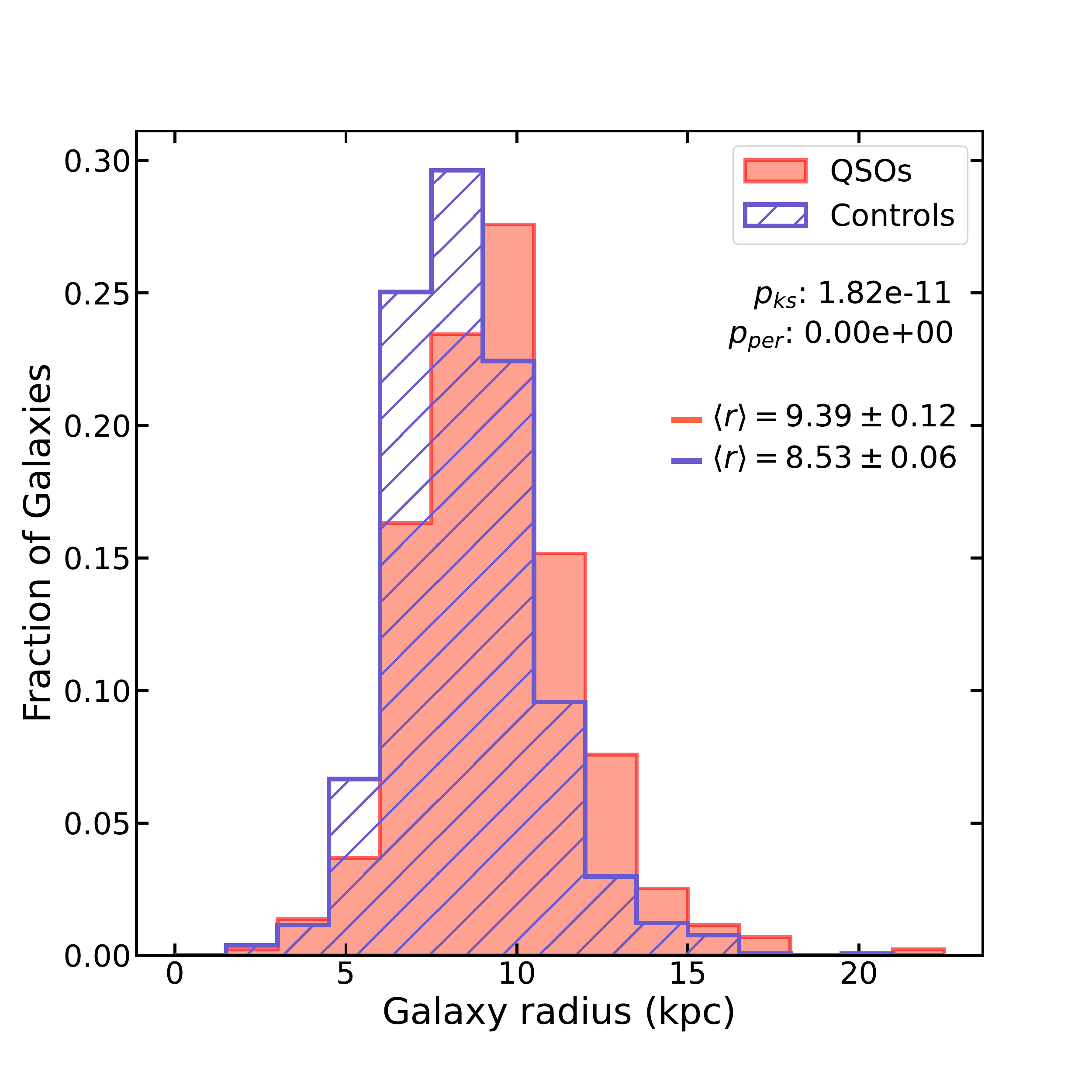}
    \caption{Histogram for the radius of the QSOs (red) and control galaxies (blue), where the p-values and median values $\langle r \rangle$ for each sample are also given.}
    \label{fig:hist_gal_radius}
\end{figure}

From the area of the galaxies we calculate the typical radius of each galaxy with $r = \sqrt{S/\pi}$. Figure \ref{fig:hist_gal_radius} shows the histogram for the radius of the QSO sample and control sample, their mean radii are respectively $\langle r \rangle = 9.39 \pm 0.12 \,$kpc and $\langle r \rangle = 8.53 \pm 0.06 \,$kpc. 

\subsubsection{Galaxy morphology}

Some recent studies have argued that QSOs are hosted predominantly in late-type galaxies \citep{2021ApJ...918...22L}. As we have not paired the QSOs and control galaxies taking into account the morphology of the galaxies, one may argue that the higher asymmetry of the QSOs could be due to the fact that they are hosted in later type galaxies than the controls. Late-type galaxies are more asymmetrical than earlier type galaxies and the higher asymmetry of QSO host galaxies could thus be due to the spiral nature. 
For example, in the classic asymmetry measurement, since it takes into account the flux distributions internal to the galaxy, spiral arms could be the cause for some of the asymmetry. 

\begin{figure}
    \centering
    \includegraphics[width=8cm]{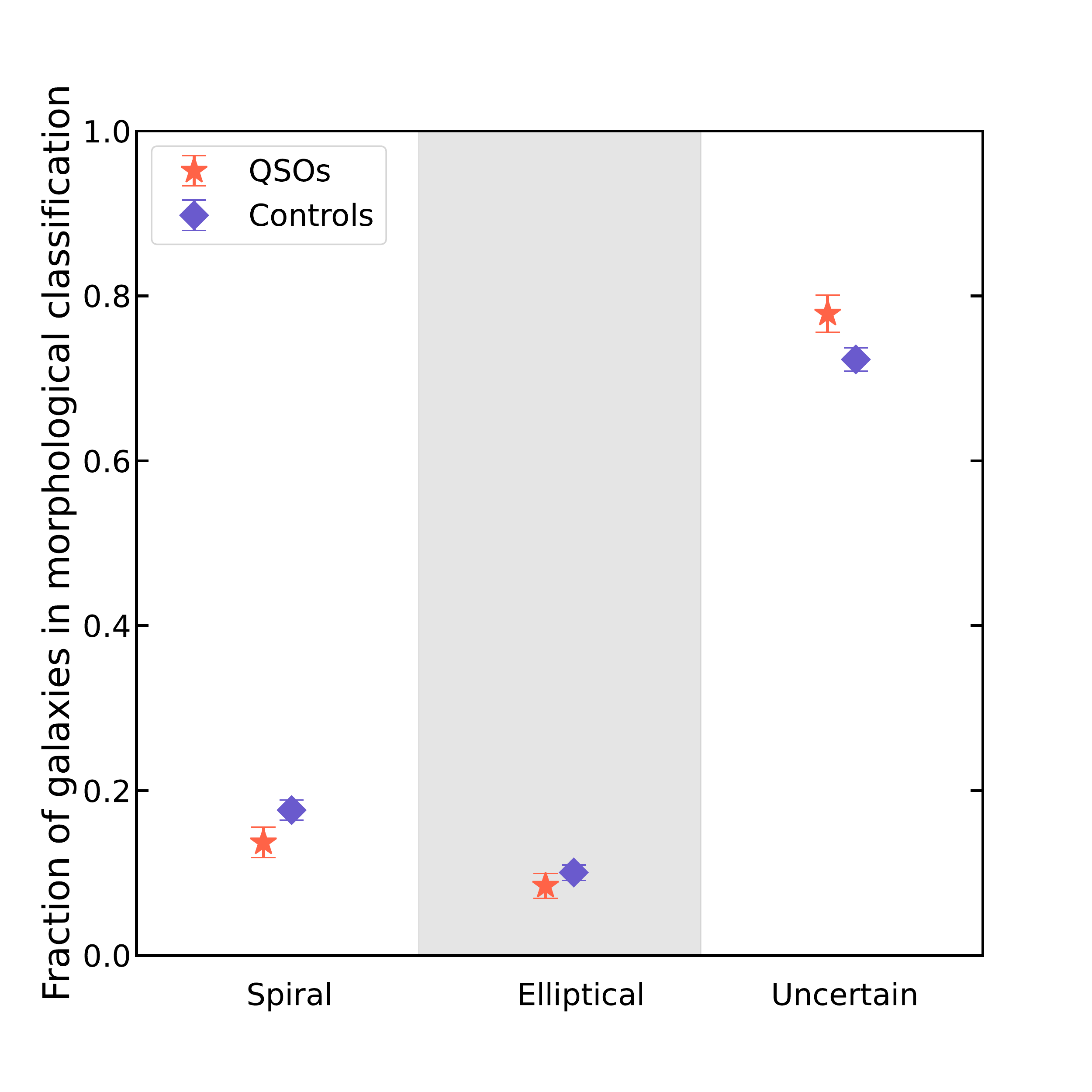}
    \caption{Fraction of galaxies according to their morphological classification in the Galaxy Zoo. QSO hosts are represented by red stars, and control galaxies by blue diamonds. The error bars are derived assuming a poisson noise. 
    }
    \label{fig:zoo}
\end{figure}

To investigate whether the higher asymmetry in the QSO sample could be a result of a difference in the morphology of the galaxies, we looked at the ``Galaxy Zoo" \citep{2011MNRAS.410..166L} classifications for the galaxies in our sample, which is compared to that of the control galaxies in Figure \ref{fig:zoo}.
As can be seen in this figure, the fraction of spiral and elliptical galaxies is very similar in the two samples. We have also investigated the possible differences in morphology dividing the sample in redshift bins and did not find differences as well in any of the redshift bins.

We thus conclude that there is no difference in the morphological classification between QSOs and controls, and we conclude that the galaxy morphology is not the cause of the higher asymmetry of the QSO relative to the controls.

\subsection{Comparison to previous studies}

We can compare our study with previous ones looking for the relation between environment and mergers in QSOs. Previous studies have indeed found that in the most luminous cases, the nuclear activity is connected to an excess of mergers, especially major mergers \citep{2012ApJ...758L..39T, 2015ApJ...806..218G, 2016ApJ...822L..32F}.

In a recent study, \citet{2021ApJ...909..124S} analysed a sample of major merging systems with projected nuclei separations between 3-15\,kpc, at $0.3<z<2.5$ and found a similar fraction of AGN in merging and non-merging systems, which they conclude indicates that mergers are inefficient in increasing AGN activity at these redshifts. The only relation found with nuclear activity in their sample was a higher AGN fraction in star-forming than in quiescent galaxies in both mergers and non-mergers. As they also mention in their discussion, some of the non-merging galaxies might be in the coalescence phase of the merger, when the SFR (star-formation rate) of the galaxy is still very high, so the correlation between SFR and AGN activity found in the non-merging galaxies could be due to a merger in a coalescence phase.

The conflicting result found in our study compared to \citet{2021ApJ...909..124S} may be attributed to the difference in redshift range used in both studies, where their sample includes high-z AGNs ($0.3<z<2.5$) and our sample focuses on low-z QSOs ($0.03<z<0.3$). This difference in redshift range may lead to variations in the physical conditions of the Universe, which can affect the evolution of galaxies and their nuclear activity.

Another possibility to explain the discrepancy between our findings and those from \citet{2021ApJ...909..124S} is the fact that our sample is restricted to QSO 2s, a minority among optically selected quasars, whose population is dominated by type 1 sources. This difference would then imply that type 1 and type 2 QSOs are not the same object seen at different sight-lines, but that they are intrinsically different. In an evolutionary scenario proposed by some authors \citep[e.g.][]{2021MNRAS.504.4445V,2022MNRAS.517.3377S}, type 2 QSOs would be in a previous stage to that of type 1 QSOs: a merger could send gas inwards and trigger nuclear activity that appears first as a type 2 QSO, in which the nuclear source -- accretion disc and broad-line region, are still obscured by the material sent inwards by the interaction or merger. Then the radiation pressure and winds from the recently formed accretion disc blows most of the obscuring material away, when the signature of the merger is already gone. The much smaller number of QSO\,2's when compared to that of QSO\,1's implies also that the QSO 2 phase is short. 

Another study finding a difference between the environment of type 1 and type 2 AGN is that of \citet{2016ApJ...832..111J} who found, for a nearby sample ($z<0.09$) of AGN and control non-active galaxies, that both type 2 AGN and normal galaxies have similar environments. The same seems to apply for type 1 AGN when one considers scales larger than $\sim$ 100 Mpc. But when smaller scales are considered, the type 2's have an excess of neighbours relative to type 1's, a result that was also found by \citet{2018MNRAS.479.2308B} for more distant AGN ($0.3<z<1.1$)
On the other hand, a more recent study of Type 1 quasars at $0.2<z<0.8$, similar to ours,   has been done using  images obtained with the Subaru Hyper Suprime-Cam \citep{2023MNRAS.521.5272T}, comparing the images of the quasars with those of a control sample. These better resolution images (than the SDSS ones used here) allowed them to separate the effect of the PSF in the images in order to investigate the host galaxy morphology. Similarly to what we have found, they have concluded that Type 1 quasars also have increased asymmetries when compared to non-active control galaxies. Although  \citet{2023MNRAS.521.5272T} do not separate their sample in type 1 and type 2, they seem to be dominated by type 1 sources, and their results, when compared to ours,  suggest that the there is not a strong difference between the close environment of type 1 and 2  quasars.

\section{Conclusions}\label{sec: concl}

We investigated a sample of 436 QSO\,2's and a matched control sample of 1308 galaxies at $0<z<0.3$, in order to quantify and compare their environment and interaction signatures as measured from SDSS-III continuum images. We quantify their number of neighbours $N$, compute their tidal strength parameter $Q$ and measure two types of asymmetry for each galaxy: classical asymmetry $A$ and shape asymmetry $A_s$. The main conclusions we obtained from this study are:  

\begin{itemize}
    \item We find a small excess in the number of neighbour galaxies of the QSO\,2's within 50\,kpc from the host only when we compare each QSO with its three controls: in 53.8\% of the QSO hosts there are more companions than in the control galaxies;
    
    \item There is no difference in the tidal strength parameter $Q$ between the QSO 2's and controls within 50\,kpc from the hosts; 
    
    \item The main difference is found in the asymmetries $A$ and $A_s$: QSO\,2 host galaxies are more asymmetrical than non-active galaxies, as measured via both parameters;
    
    \item The classical asymmetry parameter $A$, that takes into account the flux distribution, is, on average, larger than the shape asymmetry that does not contain this information, suggesting that the asymmetry is due to the presence of a disturbance that is already affecting the innermost regions of the QSO host galaxy;
    
    \item Comparing our SDSS images with HST images available for two QSO\,2 hosts with high asymmetry we conclude that the asymmetry is due to the presence of a close companion at  $\sim$5\,kpc from their nuclei;
    
    \item The area covered by the QSO hosts is larger than that of the controls, corresponding to a mean radius for the QSOs of $9.39 \pm 0.12 \,$kpc and $8.53 \pm 0.06 \,$kpc for the controls; 
    
    \item The larger area of the QSO hosts supports the presence of a close companion or disturbance in its outer region due to a merger; the fact that the close companions seen in the HST images are closer than the mean galaxy radii also supports an on-going merger; 
    
    \item The QSO's host galaxies morphological types are similar to those of their controls, ruling out the possibility that the asymmetry could be due to possible differences in the host galaxy morphology. 
    
\end{itemize}

Our main conclusion is thus that interactions do have an influence on the triggering of nuclear activity of our QSO 2's sample, that occurs only when the companion galaxy is at distances comparable to the radii of the host galaxies ($\lesssim$9\,kpc), thus taking place when they are already in the process of merging.

\section*{Data Availability}

There is no new data obtained for this paper. We have used the SDSS-III and SDSS-IV database (https://www.sdss.org/). Tables with the measurements used in the paper will be shared upon request to the corresponding author.

\bibliographystyle{mnras}
\bibliography{biblio_complete}

\label{lastpage}
\end{document}